\documentclass[aps,showpacs]{revtex4}
\begin{document}
\title{Strong gravitational force induced by static electromagnetic fields}
\author{Boyko V. Ivanov}
\email{boyko@inrne.bas.bg}
\affiliation{Institute for Nuclear Research and Nuclear Energy,\\
Tzarigradsko Shausse 72, Sofia 1784, Bulgaria}

\begin{abstract}
It is argued that static electric or magnetic fields induce
Weyl-Majumdar-Papapetrou solutions for the metric of spacetime. Their
gravitational acceleration includes a term many orders of magnitude stronger
than usual perturbative terms. It gives rise to a number of effects, which
can be detected experimentally. Four electrostatic and four magnetostatic
examples of physical set-ups with simple symmetries are proposed. The
different ways in which mass sources enter and complicate the pure
electromagnetic picture are described.
\end{abstract}

\maketitle

\section{Introduction}

In classical Newton-Maxwell physics the electromagnetic (EM) fields have no
influence upon gravity, which is generated by sources of mass. In general
relativity EM fields alter the metric of spacetime and induce a
gravitational force through their energy-momentum tensor 
\begin{equation}
T_{\;\nu }^\mu =\frac 1{4\pi }\left( F^{\mu \alpha }F_{\nu \alpha }-\frac 14%
\delta _{\;\nu }^\mu F^{\alpha \beta }F_{\alpha \beta }\right) ,  \label{one}
\end{equation}
where 
\begin{equation}
F_{\mu \nu }=\partial _\mu A_\nu -\partial _\nu A_\mu  \label{two}
\end{equation}
is the electromagnetic tensor and $A_\mu $ is the four-potential. $T_{\;\nu
}^\mu $ enters the r.h.s. of the Einstein equations 
\begin{equation}
R_{\;\nu }^\mu =\kappa T_{\;\nu }^\mu .  \label{three}
\end{equation}
We have taken into account that $T_{\;\mu }^\mu =0$. The Einstein constant
is 
\begin{equation}
\kappa =\frac{8\pi G}{c^4}=2.07\times 10^{-48}s^2/cm.g,  \label{four}
\end{equation}
where $G=6.674\times 10^{-8}cm^3/g.s^2$ is the Newton constant and $%
c=2.998\times 10^{10}cm/s$ is the speed of light. We shall use the Gauss
system (CGS) of nonrelativistic units and occasionally the international
system of practical units.

In addition, the Maxwell equations are coupled to gravity through the
covariant derivatives of $F_{\mu \nu }$%
\begin{equation}
F_{\quad ;\nu }^{\mu \nu }=\frac 1{\sqrt{-g}}\left( \sqrt{-g}F^{\mu \nu
}\right) _\nu =-\frac{4\pi }cJ_\mu ,\qquad J^\mu =\sigma cu^\mu .
\label{five}
\end{equation}
Here $g$ is the metric's determinant, usual derivatives are denoted by
subscripts, $J^\mu $ is the four-current, $u^\mu =dx^\mu /ds$ is the
four-velocity of the charged particles with charge density $\sigma $. We
shall study mainly electrovacuum solutions with $\sigma \neq 0$ only on some
surface, specifying the boundary conditions. The Einstein-Maxwell equations
(3,5) show how the EM-field leaves its imprint on the metric, which has to
satisfy the Rainich conditions \cite{one,two,three}.

The gravitational force acting on a test particle is represented by the
four-acceleration 
\begin{equation}
g_\mu =c^2\frac{du_\mu }{ds}=c^2\Gamma _{\alpha ,\mu \beta }u^\alpha u^\beta
=\frac{c^2}2g_{\alpha \beta ,\mu }u^\alpha u^\beta ,  \label{six}
\end{equation}
where $\Gamma _{\mu \beta }^\alpha $ are the Christoffel symbols. At rest $%
u^0=\left( g_{00}\right) ^{-1/2}$ and 
\begin{equation}
g_\mu =\frac{c^2}2\left( \ln g_{00}\right) _\mu  \label{seven}
\end{equation}
for an arbitrary metric $g_{\alpha \beta }$.

In this paper we investigate the problem whether EM-fields can induce strong
enough acceleration, rising above the gravimeter's threshold of $%
10^{-6}cm/s^2$ or even comparable to the mean Earth acceleration $%
g_e=980.665cm/s^2$. Eqs. (3,4) show that the metric will be very near to the
flat one without any singularities and faraway from the metric of a black
hole. The question is whether the 20 orders of magnitude supplied by $c^2/2$
in Eq. (7) are enough to lift the EM-gravitational force to that of the
Newtonian gravity of very massive bodies. In fact, we should consider the
contravariant physical (tetrad) four-vector $g^{\left( \mu \right) }=\eta
^{\mu \nu }g_{\left( \nu \right) }$, where $\eta _{\mu \nu }=diag\left(
1,-1,-1,-1\right) $, but in cartesian coordinates and for such an almost
flat metric it is indistinguishable from $g_\mu $ or $g_{\left( \mu \right)
} $ except for a sign change.

It seems natural to use perturbation theory in the harmonic gauge where 
\begin{equation}
g_{\mu \nu }=\eta _{\mu \nu }+h_{\mu \nu },\qquad \Delta h_{\mu \nu
}=-2\kappa T_{\mu \nu }  \label{eight}
\end{equation}
and the Poisson equation shows that $h_{\mu \nu }$ is extremely small, while 
$g_\mu \sim c^2\kappa =1.85\times 10^{-25}CGS$.

More concretely, one can ask what is the gravitational acceleration inside a
solenoid, the most common device for the creation of magnetic fields. There
is an exact global solution when the solenoid is infinitely long \cite{four}%
. It is comprised from the Melvin geon \cite{five,six} as an interior and
the vacuum Levi-Civita cylindrical solution \cite{seven} as an exterior. The
acceleration has only a radial component and when the metric's signature and
the scale of the magnetic field is changed correspondingly, reads inside the
solenoid 
\begin{equation}
g_{\left( r\right) }=\frac{c^2\kappa }{16\pi }\frac{rH_0^2}{A^2},\qquad A=1+%
\frac \kappa {32\pi }r^2H_0^2.  \label{nine}
\end{equation}
$H_0$ is the physical magnetic field at the central axis. The acceleration
points inwards and reaches its maximum at the solenoid's wall $r=r_1$. When
,e.g., $r_1=1m$ we have 
\begin{equation}
g_{\left( r\right) \max }=3.69\times 10^{-27}H_0^2.  \label{ten}
\end{equation}
Although the dependence on the magnetic field is quadratic, one needs $%
H_0=1.65\times 10^{10}G$ to rise above the gravimeter's threshold and $%
H_0=5.15\times 10^{14}G$ to equal $g_e$. Such strong fields may be present
on pulsars, but on Earth the static record is $33.2T$ (from HFML of the
Nijmegen University), while in pulse regime $60T$ has been obtained.

Formula (9) has been confirmed by an approximate solution to the Ernst
equation for a finite solenoid of spherical form \cite{eight}. Exact
solutions were found also for the solenoid's analogues in three \cite{nine}
and two \cite{ten} dimensions and expressions similar to Eq. (9) have been
derived. It appears that the problem has been solved and the effect is
negligible. However, there is a loophole, which is discussed in the sequel.

In Sec. II it is shown that the gravitational acceleration in
Weyl-Majumdar-Papapetrou (WMP) fields \cite{eleven,twelve,thirt,fourt}
includes a term proportional to $\sqrt{\kappa }$ and linear in the fields.
In Sec. III axially-symmetric static Einstein-Maxwell fields are revisited.
The three classes of Weyl solutions are described, as well as their relation
to the general solution and situations where they become the most general
solution. It is argued that charge and current distributions determine their
harmonic master-potential and induce them as pure electromagnetic effect
upon the spacetime metric. A general formula is given for the acceleration
and the question of hidden mass sources is investigated. In Sec. IV a short
review is given of WMP fields in the general static case, in the presence of
charged matter, pressureless (dust) or perfect fluid. The connection between
axially-symmetric static electro(magneto)vacs and stationary vacuum
solutions is elucidated.

In Sec. V we give four electrostatic examples of charged surfaces, which
induce Weyl solutions. They involve plane, spherical, spheroidal or
cylindrical symmetry. The charged plane, the Reissner-Nordstr\"om (RN)
solution and the charged cylinder are discussed from a Weyl point of view.
The peculiarities in the junction conditions for Weyl fields are pointed
out. Different physical issues, such as the force arising inside a charged
capacitor, repulsive gravity around a charged sphere and a chamber with
artificial gravity are explored. In Sec. VI we give four magnetostatic
examples, some of them being analogues to the examples in the previous
section. The first two deal with current loops, open solenoids and closed
spherical or spheroidal solenoids, which induce Weyl fields with enhanced
acceleration. The third example is the infinitely long solenoid mentioned in
the introduction. It has deep relations to Weyl fields. The genesis of Eq.
(9) is described in detail. The fourth example is the gravitational field of
a line current, which is similar to Weyl fields with cylindrical symmetry.
In Sec. VII the main results about WMP fields are summarised into 19 points.
The last section contains discussion of their role in physics and some
historical remarks.

\section{Root gravity}

Let us assume that the metric and the EM-fields do not depend on time. In
this stationary case let us further simplify the problem by setting $A_\mu
=\left( \bar \phi ,0,0,0\right) $. There is just an electric field 
\begin{equation}
E_\mu =F_{0\mu }=-\bar \phi _\mu .  \label{eleven}
\end{equation}
Obviously, $T_{\;\nu }^\mu $ from Eq. (1) contains only quadratic terms in $%
\bar \phi _\mu $. This allows to hide $\kappa $ from Eq. (3) by normalizing
the electric potential to a dimensionless quantity 
\begin{equation}
\phi =\sqrt{\frac \kappa {8\pi }}\bar \phi .  \label{twelve}
\end{equation}
The factor $8\pi $ is chosen for future convenience. We shall see that this
is a much more elegant way to get rid of the constants in the
Einstein-Maxwell equations than the choice of relativistic units $c=1$ and $%
G=1$ ($8\pi G=1$ sometimes).

Imagine now that in some exact solution $g_\mu $ is proportional to the
electric field, contrary to the quadratic dependence in Eq. (9) 
\begin{equation}
g_\mu =ac^2\phi _\mu =ac^2\sqrt{\frac \kappa {8\pi }}\bar \phi _\mu ,
\label{thirt}
\end{equation}
where $a$ is some slowly changing function of order $O\left( 1\right) $. For
example, when $a=a\left( g_{00}\right) $ then Eqs. (7,13) lead to the
functional dependence $f\equiv g_{00}=F\left( \phi \right) $. Let us further
assume that the spacetime is static. Then the above functional dependence
has the unique form \cite{eleven,twelve,thirt} 
\begin{equation}
f=1+B\phi +\phi ^2.  \label{fourt}
\end{equation}
In the axially-symmetric case Eq. (14) was found by Weyl already in 1917 
\cite{fourt} and such solutions are known as Weyl fields. The potential in
Eq. (12) is very small everywhere and naturally goes to zero at infinity.
Then asymptotic flatness fixes the first term which is otherwise an
arbitrary constant. It is also fixed by the requirement to go back to
Minkowski spacetime when $\phi =0$ since we are studying the EM-effect on
gravity with no masses present. Hence, we should not make any gauge
transformations $\phi \rightarrow \phi +C$, which can eliminate the
arbitrary constant $B$ but spoil the behaviour of the metric. Later we will
give arguments that the typical value of $B$ is $2$ and not zero, so that
the linear term in Eq. (14) is really present.

Thus in Weyl-Majumdar-Papapetrou fields we have 
\begin{equation}
g_\mu =c^2f^{-1}\left( \frac B2\sqrt{\frac \kappa {8\pi }}\bar \phi _\mu +%
\frac \kappa {8\pi }\bar \phi \bar \phi _\mu \right) .  \label{fift}
\end{equation}
The first term is of the type given in Eq. (13), the second resembles the
expression in Eq. (9). Let us note that 
\begin{equation}
c^2\sqrt{\frac \kappa {8\pi }}=\sqrt{G}=2.58\times 10^{-4},\qquad c^2\frac 
\kappa {8\pi }=\frac G{c^2}=7.37\times 10^{-27}.  \label{sixt}
\end{equation}
Due to the square root, the first coefficient is $10^{23}$ times bigger than
the second. We shall call gravitational fields, which have acceleration
terms $\sim \sqrt{\kappa }$, root gravity. The WMP fields are an example,
but there are others too. Thus general relativity has a Newtonian limit in
the case of mass sources, where $g_\mu \sim G$ and a Maxwellian limit in the
case of EM-sources, where $g_\mu \sim \sqrt{G}$.

Provided that $B\approx 2$ our search for a strong gravitational
acceleration induced by EM-fields doesn't seem so doomed as in the
introduction. Electrostatic generators can create potential differences of
six million volts or 
\begin{equation}
\bar \phi _{\max }=2\times 10^4CGS.  \label{sevent}
\end{equation}
If it were applied to a capacitor with distance of $1cm$ between the plates,
the electric field will be of the same order. It compensates the root
coefficient in Eq. (15) and we get acceleration of about $1cm/s^2$, which is
perfectly measurable. Static magnetic fields create the same gravitational
effects as static electric fields \cite{fift,sixt,sevent}, so a field of $%
33.2T$ may induce in principle acceleration of about $10cm/s^2$. One needs
just two orders more to counter Earth's gravity. These effects are much
stronger than any other known general relativistic effects, including
gravitational waves and gravitomagnetism, which are currently under
intensive study. They may be produced in a finite region in a laboratory if
we learn how to create WMP fields.

Up to now the only sources of gravitation have been masses (preferably in
enormous quantities), therefore, it is worth to compare them to root gravity
in equal conditions. Let us take a ball of radius $R$, mass density $\mu $
and total mass $M$. The acceleration at its surface is 
\begin{equation}
g_R=\frac{GM}{R^2}=\frac{4\pi }3G\mu R=2.8\times 10^{-7}\mu R.
\label{eightt}
\end{equation}
Let the substance of the ball be Osmium, which possesses one of the biggest
densities, $\mu =22.48g/cm^3$. For a laboratory set-up $R\approx 10^2cm$.
This yields $6.44\times 10^{-4}cm/s^2$, which is much less than the root
gravity results. Comparing Eqs. (14) with $B=2$ and (18) we get the
effective EM-field necessary to induce the same gravitational acceleration 
\begin{equation}
E_{eff}=1.08\times 10^{-3}\mu R.  \label{ninet}
\end{equation}
In other words, it is thousand times more effective than a mass source. Root
gravity is about $10^{18}$ times stronger than usual perturbative gravity
arising from Eq. (8), than the exact solution in Eq. (9) and than the second
term in Eq. (15). Therefore, in the next sections we revisit the WMP fields,
putting the emphasis on their physical applicability.

\section{Weyl fields revisited}

Let us start with the axially-symmetric static metric 
\begin{equation}
ds^2=f\left( dx^0\right) ^2-f^{-1}\left[ e^{2k}\left( dr^2+dz^2\right)
+r^2d\varphi ^2\right] ,  \label{twenty}
\end{equation}
where $x^0=ct$, $x^1=\varphi ,$ $x^2=r,$ $x^3=z$ are cylindrical
coordinates, $f=e^{2u}$ and $u$ is the first, while $k$ is the second
gravitational potential. Both of them depend only on $r$ and $z$. Let $A_\mu
=\left( \bar \phi ,\bar \chi ,0,0\right) $ where $\bar \chi $ is the true
magnetic potential. Following Tauber \cite{eightt} we introduce the
auxiliary potential $\bar \lambda $%
\begin{equation}
\lambda _r=\frac fr\chi _z,\qquad \lambda _z=-\frac fr\chi _r,  \label{twone}
\end{equation}
so that 
\begin{equation}
F^{\varphi r}=\frac{fe^{-2k}}r\bar \lambda _z,\qquad F^{z\varphi }=\frac{%
fe^{-2k}}r\bar \lambda _r  \label{twtwo}
\end{equation}
describe the axial and the radial components of the magnetic field. For the
electric field one has 
\begin{equation}
E_r=F_{0r}=-\bar \phi _r,\qquad E_z=F_{0z}=-\bar \phi _z.  \label{twthree}
\end{equation}
The field equations read 
\begin{equation}
\Delta u=e^{-2u}\left( \phi _r^2+\phi _z^2+\lambda _r^2+\lambda _z^2\right) ,
\label{twfour}
\end{equation}
\begin{equation}
\Delta \phi =2\left( u_r\phi _r+u_z\phi _z\right) ,  \label{twfive}
\end{equation}
\begin{equation}
\Delta \lambda =2\left( u_r\lambda _r+u_z\lambda _z\right) ,  \label{twsix}
\end{equation}
\begin{equation}
\phi _r\lambda _z=\phi _z\lambda _r,  \label{twseven}
\end{equation}
\begin{equation}
\frac{k_r}r=u_r^2-u_z^2-e^{-2u}\left( \phi _r^2-\phi _z^2+\lambda
_r^2-\lambda _z^2\right) ,  \label{tweight}
\end{equation}
\begin{equation}
\frac{k_z}r=2u_ru_z-2e^{-2u}\left( \phi _r\phi _z+\lambda _r\lambda
_z\right) ,  \label{twnine}
\end{equation}
\begin{equation}
k_{rr}+k_{zz}=\Delta u-\left( u_r^2+u_z^2\right) ,  \label{thirty}
\end{equation}
where $\Delta =\partial _{rr}+\partial _{zz}+\partial _r/r$. We have used
the definition given in Eq. (12) and a similar one for $\lambda $. Using Eq.
(27) one can prove that $\lambda =\lambda \left( \phi \right) $ and the
dependence is linear \cite{fift,sixt}. This result holds also for a general
static metric \cite{sevent}. It is enough to engage just an electric field,
there being a trivial magnetovac analogue to every electrovac solution. Eqs.
(24-27) reduce to \cite{ninet} 
\begin{equation}
\Delta u=e^{-2u}\left( \phi _r^2+\phi _z^2\right) ,\qquad \Delta \phi
=2\left( u_r\phi _r+u_z\phi _z\right) ,  \label{thone}
\end{equation}
which determine $\phi $ and $f$. Eqs. (28,29) become 
\begin{equation}
\frac{k_r}r=u_r^2-u_z^2-e^{-2u}\left( \phi _r^2-\phi _z^2\right) ,\qquad 
\frac{k_z}r=2u_ru_z-2e^{-2u}\phi _r\phi _z  \label{thtwo}
\end{equation}
and determine $k$ by integration, while Eq.(30) holds identically and is
redundant. When $\phi =0$, Eq. (31) becomes the Laplace equation for $u$ and
Eq. (32) gives $k\left( u\right) $. These are the axially-symmetric vacuum
equations, also discovered by Weyl.

Now let us make the assumption that the gravitational and the electric
potential have the same equipotential surfaces, $f=f\left( \phi \right) $.
Eq. (31) yields 
\begin{equation}
\left( f_{\phi \phi }-2\right) \left( \phi _r^2+\phi _z^2\right) =0
\label{ththree}
\end{equation}
and that's how the quadratic relation (14) appears. Replacing Eq. (14) in
Eq. (31) one comes to an equation for $\phi $%
\begin{equation}
\Delta \phi =\frac{B+2\phi }{1+B\phi +\phi ^2}\left( \phi _r^2+\phi
_z^2\right) .  \label{thfour}
\end{equation}
We put for definiteness $B\geq 0$. Eqs. (14,34) with $B\leq 0$ are obtained
by changing the sign of $\phi $. The general solution of Eq. (34) is not
known. However, let us make one more assumption, that $\phi $ depends on $%
r,z $ through some function $\psi \left( r,z\right) $. Eq. (34) becomes 
\begin{equation}
\frac{\phi _{\psi \psi }}{\phi _\psi }-\frac{\left( B+2\phi \right) \phi
_\psi }{1+B\phi +\phi ^2}=-\frac{\Delta \psi }{\psi _r^2+\psi _z^2}.
\label{thfive}
\end{equation}
If $\psi $ satisfies the Laplace equation $\Delta \psi =0$, $\phi \left(
\psi ,B\right) $ is determined implicitly \cite{fourt} from 
\begin{equation}
\psi =\int \frac{d\phi }{1+B\phi +\phi ^2}.  \label{thsix}
\end{equation}
A very important equality follows 
\begin{equation}
\phi _i=f\psi _i,\qquad \bar \phi _i=f\left( \phi \right) \bar \psi _i,
\label{thseven}
\end{equation}
where $i=r,z$. Eq. (32) becomes 
\begin{equation}
k_r=\frac D4r\left( \psi _r^2-\psi _z^2\right) ,\qquad k_z=\frac D2r\psi
_r\psi _z,  \label{theight}
\end{equation}
where $D=B^2-4$. Obviously, $k\sim \kappa $ always, making it much smaller
than $u$.

Thus in Weyl electrovac solutions the harmonic master potential $\psi $
determines the electric and the gravitational fields like $u$ does this in
the vacuum case. One may go further and find a relation between $\psi $ and $%
u$, transforming Weyl electrovacs into Weyl vacuum solutions, although usual
transformations work the other way round \cite{three}. In particular
solutions $\phi $ is usually proportional to the charge, $\phi =q\tilde \phi 
$. Eq. (36) shows that $\psi =q\tilde \psi $ where $\tilde \psi $ is
harmonic and finite when the electric field is turned off by $q\rightarrow 0$%
. In this limit, when $B$ does not depend on $q$, we have $f\rightarrow 1$
from Eq. (14) and $k\rightarrow 0$ from Eq. (38). Trivial flat spacetime is
the result. However, if $B=\tilde B/q$ then $f\rightarrow 1+\tilde B\tilde 
\phi $, $f_i\rightarrow \tilde B\tilde \phi _i$ and from Eq. (37) it follows
that $u=\tilde B\tilde \psi /2$ and is harmonic. Eq. (38) then gives the
vacuum expression for $k$. Hence, we obtain a Weyl vacuum solution with the
same (up to a constant) harmonic function $\psi $. A mass term has appeared
out of the vanishing charge.

Let us go back to the electrovac problem. One can add a constant $\psi _0$
to $\psi $ in order to satisfy the conditions $\psi \rightarrow
0,f\rightarrow 1,\phi \rightarrow 0$ at infinity or when the electric field
is turned off. The integral in Eq. (36) can be analytically evaluated and
the dependence $\phi \left( \psi ,B\right) $ made explicit. There are three
cases, according to the sign of $D$. The simplest one is $D=0$ ($B=2$). Then 
$f$ becomes a perfect square and $\psi _0=-1$, 
\begin{equation}
\phi =-1-\frac 1{\psi +\psi _0}=\frac \psi {1-\psi },\qquad f=\left( 1-\psi
\right) ^{-2}.  \label{thnine}
\end{equation}

When $D<0$ Eq. (38) gives $-D<4$ and trigonometric functions appear 
\begin{equation}
\phi =-\frac B2+\frac{\sqrt{-D}}2\tan \frac{\sqrt{-D}}2\left( \psi +\psi
_0\right) ,\qquad \psi _0=\frac 2{\sqrt{-D}}\arctan \frac B{\sqrt{-D}},
\label{forty}
\end{equation}
\begin{equation}
f=-\frac D{4\cos ^2\frac{\sqrt{-D}}2\left( \psi +\psi _0\right) }=\left(
\cos \frac{\sqrt{-D}}2\psi -\frac B{\sqrt{-D}}\sin \frac{\sqrt{-D}}2\psi
\right) ^{-2}.  \label{foone}
\end{equation}
When $\psi _0\equiv 0$ these formulas coincide with the Bonnor's ones \cite
{ninet}.

Finally, when $D>0$ there exist two expressions for the integral, one as a
logarithm, the other in hyperbolic functions. They lead to 
\begin{equation}
\phi =-\frac B2-\frac{\sqrt{D}}2\coth \frac{\sqrt{D}}2\left( \psi +\psi
_0\right) =\frac{2\left( e^{\sqrt{D}\psi }-1\right) }{B+\sqrt{D}-\left( B-%
\sqrt{D}\right) e^{\sqrt{D}\psi }},  \label{fotwo}
\end{equation}
\begin{equation}
f=\frac D{4\sinh ^2\frac{\sqrt{D}}2\left( \psi +\psi _0\right) }=\left(
\cosh \frac{\sqrt{D}}2\psi -\frac B{\sqrt{D}}\sinh \frac{\sqrt{D}}2\psi
\right) ^{-2},  \label{fothree}
\end{equation}
\begin{equation}
e^{\sqrt{D}\psi _0}=\frac{B-\sqrt{D}}{B+\sqrt{D}}.  \label{fofour}
\end{equation}

According to Bonnor (who does not introduce $\psi _0$) the expressions for $%
f $ and $\phi $ in the case $D>0$ are obtained from Eq. (40) by continuation
of $\sqrt{-D}$ to imaginary values and consequently $\tan \rightarrow \tanh $%
, $\cos \rightarrow \cosh $. This, however, holds when $\left( 2\phi
+B\right) ^2<D$ , i.e., $4f<0$, which is unphysical. In the physical case we
must also do the replacement $\tanh \rightarrow \coth $, $\cosh \rightarrow
\sinh $. The above discussion shows that the point $B=2$ has a privileged
position, unlike $B=0$.

The Weyl solutions were derived with two assumptions imposed on the system: $%
f=f\left( \phi \right) $; $\phi =\phi \left( \psi \right) $, $\Delta \psi =0$%
. They are particular solutions of Eqs. (31,32). However, when the symmetry
is stronger than axial and the fields depend on just one coordinate $x$ (not
necessarily cylindrical, but a function of $r,z$), they comprise the general
solution. For then $f\left( x\right) $ and $\phi \left( x\right) $ obviously
are functionally related and one can always find $X\left( x\right) $ so that 
$\Delta X\left( x\right) =0$. Taking $\psi =X\left( x\right) $ and
expressing $f$ and $\phi $ as functions of $X$, leads inevitably to the Weyl
solutions. In the case of plane symmetry $x=z$, $X=x$. Cylindrical symmetry
gives $x=r$, $X=\ln x$. Spherical symmetry has $x=\sqrt{r^2+z^2}$, $X=1/x$.

We have described the advantages of WMP solutions in inducing a powerful
gravitational force. The natural questions appear: is it possible to create
such solutions in a laboratory? What kind of charged sources should we take?
Point \cite{twenty} and line \cite{twone} sources lead to singularities and
other problems. Therefore, we take a charged closed rotationally-symmetric
surface with invariant density of the surface charge $\sigma _s$. The
electrostatic theorem of Gauss has a generalization in general relativity 
\cite{thirt,twtwo}. Integrating the r.h.s. of Eq. (5) one obtains the total
charge contained in some volume 
\begin{equation}
e=\frac 1c\int J^0\sqrt{-g}d^3S=\int \sigma _3d^3S,\quad \sigma _3=\sigma 
\sqrt{-g^{\left( 3\right) }},  \label{fofive}
\end{equation}
where $\sigma _3$ is the three-dimensional invariant density and $g^{\left(
3\right) }$ is the determinant of the space part of the metric. When the
charge is attached to a surface, one should use $\sigma _s$ instead.
Integration of the l.h.s. of Eq. (5) leads to a relation between $e$ and the
electric flux through a closed surface $S$, encompassing the charged volume

\begin{equation}
4\pi e=\int_S\left[ F^{01}\frac{\partial \left( x_2,x_3\right) }{\partial
\left( u,v\right) }+F^{02}\frac{\partial \left( x_3,x_1\right) }{\partial
\left( u,v\right) }+F^{03}\frac{\partial \left( x_1,x_2\right) }{\partial
\left( u,v\right) }\right] \sqrt{-g}dudv.  \label{fosix}
\end{equation}
For Weyl solutions 
\begin{equation}
F^{0i}\sqrt{-g}=-r\bar \psi _i,  \label{foseven}
\end{equation}
which is the result for flat spacetime. Hence, Eq. (46) becomes the Gauss
theorem in classical electrostatics, but with $\phi $ replaced by $\psi $,
which satisfies the Laplace equation. This fact was already stressed by
Bonnor \cite{five,ninet} but in view of its importance we have discussed it
again. Following a well-known procedure, we obtain a boundary condition on $%
S $ for the jump of the normal component $\bar \psi _n$: 
\begin{equation}
-\bar \psi _n|_{-}^{+}=4\pi \bar \sigma _s.  \label{foeight}
\end{equation}
When $\psi $ is given on $S$, there are two well-defined Dirichlet boundary
problems and it may be continued as harmonic function inside and outside $S$%
. If $\alpha \leq \psi _s\leq \beta $, these inequalities hold for $\psi $
throughout space and it will be regular. Then $f$, $k$ and $\phi $ are found
from $\psi $ in a manner already explained. The jump of $\psi _n$ at $S$
determines the source $\sigma _s$. The inverse is also true. When $\sigma _s$
is given, there is a unique global $\psi $, satisfying Eq. (48).
Consequently, for any distribution of charges on $S$ one can find the
electric and gravitational fields they induce. The same can be done when $S$
is infinite and / or not closed, but singularities may creep into the
solutions.

Finally, by replacing Eq. (37) into Eq. (15) and expressing $f=f\left( \psi
\right) $ we obtain 
\begin{equation}
g_i=\frac 12\sqrt{G\left( D+4f\right) }\bar \psi _i=\sqrt{Gf}\bar \psi
_i|_{B=2}.  \label{fonine}
\end{equation}
For realistic EM-fields $f$ is very close to one and this Maxwellian effect
is the only one to be observed. Typical Einsteinian effects like light
bending, gravitational redshift, time delay, changes in lengths are not
enhanced by $c^2$ and are negligible.

Some questions immediately arise. Why do we get a solution for an arbitrary $%
\sigma _s$ when the Weyl fields are not the most general solutions of Eq.
(31)? Why $\phi ,f,k$ depend not alone on $\psi $ but also on the constant $%
B $? How is its value determined, can we increase it, to enhance the effect
of root gravity? In order to answer them we must return to the starting
point, Eq.(3). Even when the EM-field is absent, Eq. (3) still has a number
of non-trivial vacuum solutions, including e.g. gravitational waves. The
situation is similar to classical electrodynamics without sources.
Non-trivial solutions exist, but have to be time-dependent. These are the
well-known electromagnetic waves. General relativity is a highly non-linear
theory and vacuum solutions exist also in the static case. Their sources are
well-hidden masses and even today there is a gap between the mathematical
derivation of solutions \cite{three} and their physical interpretation \cite
{twthree}. When EM-fields are turned on, these parasitic masses do not
disappear and obscure the pure effect of electromagnetism on gravity. Let us
try to get rid of them, step by step. First of all, the metric should
inherit the symmetry of EM-fields. Let us confine again ourselves to axial
symmetry. There are a lot of generation techniques, which produce non-Weyl
solutions of Eq. (31). One of them, ''coordinate modelling'', adapts the
coordinates after the equipotential surfaces of the electric field \cite
{twfour,twfive,twsix} and includes in a natural way the set of Weyl
solutions. Most of the methods (see Ref. \cite{three}, Ch.34), however,
start from the reformulation of Eq. (31) in terms of the Ernst potential $%
E=f-\phi ^2$ \cite{twseven}, which is real in the absence of rotation, 
\begin{equation}
f\Delta E=\nabla f\nabla E,\quad f\Delta \phi =\nabla f\nabla \phi .
\label{fifty}
\end{equation}
The general solution of the Ernst equations can be found when the behaviour
of $E\left( z\right) $ and $f\left( z\right) $ on the axis is given. It
determines the multipole structure and is useful in astrophysics for
modelling the gravitational field of stars. The presence of masses is
welcomed, since they give the most substantial gravitational effect,
followed by rotation (it can be incorporated into the formalism), magnetic
fields and electric charge at the last place. The Ernst equation is much
more difficult than the Laplace one and Dirichlet boundary value problems
for it were discussed only recently \cite{tweight}. On the other side, a
harmonic function may be easily restored from its values on the axis $\psi
\left( z\right) $ \cite{twnine,thirty} 
\begin{equation}
\psi \left( r,z\right) =\frac 1\pi \int_0^\pi \psi \left( z+ir\cos \theta
\right) d\theta .  \label{fione}
\end{equation}
This real expression was used in general relativity for axially-symmetric
static vacuum solutions \cite{thone}, but it is not difficult to adapt it to
Weyl electrovacs too. In the general solution $f\left( z\right) $ is not
correlated with $\phi \left( z\right) $ and can be arbitrary, even when $%
\phi \left( z\right) $ vanishes, signalling the presence of masses. In the
Weyl solution the metric goes to flat Minkowski spacetime when $\psi \left(
z\right) $ vanishes. Thus the general solution expands over the Weyl one by
the addition of masses. Probably the same is true for the solutions of Eq.
(34), which also contain root gravity terms, since Eq. (14) is satisfied.
The reason is that Weyl solutions form a complete system, covering the
effect of any charge distribution and more general solutions can include the
only other source of gravitation. More precisely, Weyl fields form an
overcomplete system due to $B$ which is not fixed by $\sigma _s$. In the
case of plane, spherical, spheroidal or cylindrical symmetry they coincide
with the most general solution of the Ernst equation (50). It is quite
improbable that the unwanted masses should disappear exactly in these cases,
so the only way to show their presence is through the value of $B$.
Curiously, in a recent paper \cite{thtwo} it is asserted that the sphere,
rod and plane are the only Weyl fields with geodesic lines of force and are
algebraically special of Petrov type D. A logical step is to accept that $%
\psi $ plays the role of $\phi $ in any situation in electrostatics, not
only for charged surfaces. Hence, when the gravitation created by charges is
taken into account, it seems that Weyl fields generalize the solution to
classical electrostatic problems. The physical electric field is found with
the help of Eq. (37) 
\begin{equation}
E_{\left( i\right) }=-\left( g_{00}g_{ii}\right) ^{-1/2}\bar \phi _i=-fe^{-k}%
\bar \psi _i.  \label{fitwo}
\end{equation}
Now, since $f,k$ are extremely close to one and zero respectively, one can
do a perturbation theory around the exact Weyl solutions and set $E_{\left(
i\right) }\approx -\bar \psi _i$. In the present case WMP fields are similar
to instantons, monopoles, solitons and other non-perturbative exact
solutions in quantum field theory. With high precision all electrostatic
formulae hold also in the Einstein-Maxwell ''already unified theory''. The
only new effect is the appearance of an electromagnetically induced
gravitational acceleration, which reads from Eq. (49), again with high
precision, 
\begin{equation}
g_i=-\frac B2\sqrt{G}E_i.  \label{fithree}
\end{equation}
We shall give some arguments in the following that $B=2$ when unbiased by
parasitic masses. Therefore, as already explained, measurable $g_i$ are
present from the already available electric and magnetic fields. One can
reach $g_e$ when $E_i=1.14\times 10^9V/cm=3.8\times 10^6CGS$ or $%
H_i=380T=3.8\times 10^6G.$ The lines of acceleration follow the electric
field lines. Test particles will stay in equilibrium if they are charged and
the relation between their mass $m$ and charge $e$ is 
\begin{equation}
\left| e\right| =\sqrt{G}m.  \label{fifour}
\end{equation}

Root gravity has some peculiar features. Changing the direction of $E_i$ one
changes the direction of $g_i$ and when it points upwards with respect to
the Earth's surface one has ''anti-gravity''. This is true because in our
perturbation theory accelerations from electric fields and masses like the
Earth or laboratory masses are added as usual vectors. The exact Weyl
solution is necessary to clarify the gravitational induction in a laboratory
set-up of finite size in space where $E_i$ is present. Although we have used
a long-range interaction to induce another long-range interaction, in
reality static EM-fields are always confined. Eq. (53) shows that putting a
Faraday cage on $E_i$ does the same on $g_i$ and confines root gravity too.
It is understandable that when non-mass sources of gravitation are applied,
the appearance of monopole terms (usually considered as mass terms) is not
obligatory. Their existence is usually based on the Whittaker's theorem \cite
{twtwo}, which demonstrates the influence of some combination of the $T_{\mu
\nu }$ components (called gravitational mass) upon the gravitational
acceleration. The relation is given in terms of surface and volume
integrals, appearing when the time-time component of the Einstein equations
(1) is integrated. In this way it is not something separate and additional
to them, but a consequence that can't contradict the conclusions following
from them. Concretely, for Weyl fields this theorem is just the Gauss
theorem for $grad\;u$. In the case of electromagnetic sources the
''gravitational mass'' is in fact some kind of ''energy'', inducing
gravitational acceleration not necessarily with a monopole term. We avoid
arguments based on the energy of the gravitational field because its density
is not a tensor and there are at least five energy-momentum complexes \cite
{ththree}, each with its own merits. Of course, some small mass term will
always exist, due to the mass of the surface $S$. However, it will be of
usual perturbative nature, many orders of magnitude smaller than root
gravity. 

Let us turn now to the case of magnetostatics. As was mentioned before, the
analogue of $\phi $ is $\lambda $. It should be replaced in Eqs.
(14,15,31,37,39,40,42). The analogue of Eq. (46) vanishes because there are
no magnetic charges. One should take a closed surface with surface current.
In an axially-symmetric problem it has just one component, $J_\varphi $. A
Weyl magnetostatic solution was given for the first time by Papapetrou in
1947 \cite{twelve}. The analogy with electrostatics was investigated by
Bonnor \cite{five,ninet} who showed that $\psi $ is equivalent to the scalar
magnetic potential. Then Eq. (48) should give the jump of the tangential to
the surface component $H_t$ which follows classically from $\bar \psi $ 
\begin{equation}
H_t|_{-}^{+}=\frac{4\pi }cJ_\varphi  \label{fifive}
\end{equation}
and is perpendicular to $J_\varphi $. Eqs. (52,53) still hold with $%
E_{\left( i\right) }\rightarrow H_{\left( i\right) }$, so that the
gravitational effects of static magnetic fields mirror that in
electrostatics. One can also imagine a vector potential, corresponding to $%
\psi $, which is more convenient in classical magnetostatics. However, in
order to find the metric, the scalar potential $\psi $ for each classical
problem should be calculated too. The fields mainly of linear sources, like
a current loop \cite{twone}, and disks \cite{thfour,thfive} have been
examined without noticing the presence of root gravity.

\section{Connections, generalizations and analogies}

It is well known that in vacuum spacetimes with isometries any Killing
vector may be considered as an electromagnetic potential and satisfies the
Maxwell equations in this background \cite{thsix}. This Papapetrou field is
used to find EM solutions from vacuum ones \cite{thseven}. In the
axially-symmetric case this leads to dependence between $\phi $ and $f$ and
one comes to solutions of Eq. (34), which contain root gravity terms because
of Eq. (14). Even a Weyl solution was derived recently, starting from the
vacuum $\gamma $-metric \cite{theight}, although this was not stated.

In the general stationary case the interval reads 
\begin{equation}
ds^2=f\left( dx^0+\omega _adx^a\right) -f^{-1}\gamma _{ab}dx^adx^b,
\label{fisix}
\end{equation}
where $\omega _a$ is the gravitomagnetic potential ($a=1,2,3$) and $\gamma
_{ab}$ is the three-dimensional metric. In the general static case $\omega
_a=0$ and in its electrostatic subcase the Einstein-Maxwell equations read 
\begin{equation}
\Delta u=e^{-2u}\nabla \phi \nabla \phi ,\quad \Delta \phi =2\nabla u\nabla
\phi ,  \label{fiseven}
\end{equation}
\begin{equation}
R_{ab}^{\left( 3\right) }=2u_au_b-2e^{-2u}\phi _a\phi _b,  \label{fieight}
\end{equation}
where $F_{0a}=-\bar \phi _a$. In magnetostatics $\phi \rightarrow \lambda $
the latter being defined by 
\begin{equation}
F^{ab}=\left( -g\right) ^{-1/2}\varepsilon ^{abc}\bar \lambda _c.
\label{finine}
\end{equation}
This equation generalizes Eq. (22). The gradients, the Laplacian and the
three-dimensional Ricci tensor are with respect to the metric $\gamma _{ab}$%
. Eqs. (57,58) generalize Eqs. (31,32). However, $\gamma _{ab}$ also enters
Eq. (57), making all equations interconnected and the system very difficult
to deal with. In the special case when Eq. (14) holds with $B=2$, $\gamma
_{ab}$ becomes flat and Eq. (57) can be solved, since it decouples from Eq.
(58). The result is Eq. (39) with a harmonic $\psi \left( \varphi
,r,z\right) $. Usually one takes $1-\psi =U$ and $\phi =U^{-1}$, obtaining
the already mentioned Majumdar-Papapetrou solutions from 1947 \cite
{eleven,twelve}, which are conformastatic. Some years later Ehlers \cite
{thnine,forty} gave transformations to derive such fields from vacuum ones.
These are the only other papers on the subject written in non-relativistic
units. Similar transformations were given by Bonnor \cite{foone} and with
the help of the TWS method \cite{fotwo,fothree}. The latter was applied to
scalar fields \cite{fotwo} and to the RN solution \cite{fothree,fofour}.

When there is no space symmetry present, the simplest harmonic function in
cartesian coordinates is 
\begin{equation}
U\left( x,y,z\right) =1+\sum\limits_i\frac{Gm_i}{c^2r_i},\quad r_i=\left[
\left( x-x_i\right) ^2+\left( y-y_i\right) ^2+\left( z-z_i\right) ^2\right]
^{1/2}.  \label{sixty}
\end{equation}

It can be shown that the sources are point monopoles with masses $m_i$ and
charges $e_i$ \cite{thirt,twenty}, connected by Eq. (54), which is true also
in the Newtonian theory. It ensures the equilibrium between the electric and
gravitational forces among the sources. Such multi-black-hole solutions
satisfy certain uniqueness theorems \cite{fofive,fosix} and possess
axially-symmetric reduction with the points staying on the $z$-axis \cite
{foseven}. They have been generalized to expanding cosmological solutions 
\cite{foeight,fonine} and the moving charged masses do not radiate
electromagnetic or gravitational waves \cite{fifty}.

Solutions based on Eq. (60) are electrovacuum solutions and have
singularities at the point sources. Increasing their number to infinity we
expect to obtain charged dust in equilibrium. However, for this purpose one
must add to the r.h.s. of Eq. (3) $T^{\mu \nu }=\mu c^2u^\mu u^\nu $ and
introduce the current in Eq. (5). In this way charged dust with mass density 
$\mu $ and charge density $\sigma $ is described. In the general static case
one can impose the Newtonian equilibrium condition 
\begin{equation}
\pm \sigma =\sqrt{G}\mu ,  \label{sione}
\end{equation}
which is the density analogue of Eq. (54), but here it holds for the sources
of the field, not for a test-particle. Then one gets \cite{fione} 
\begin{equation}
f=\left( C+\phi \right) ^2,  \label{sitwo}
\end{equation}
where $C$ is some constant. Unfortunately, charged dust clouds are interior
solutions and we cannot use asymptotic flatness to set $C=1$. Neither can we
turn off the electric field putting $\sigma =0$ for then $\mu =0$. Neutral
dust cannot be in static equilibrium - it collapses. Therefore, $C$ is
usually set to zero by a gauge transformation and then there are no root
gravity terms. The situation is not clear, but the perfect square in Eq.
(62) and the possibility of conformastatic MP solutions speak in favour of $%
B=2$ and consequently, $D=k=0$ in Eq. (38). Like in the multi-black-hole
case one puts $\phi =U^{-1}$, $f=U^{-2}$, but $U$ is not harmonic; it
satisfies the non-linear equation \cite{eleven,fione} 
\begin{equation}
\Delta U=\pm \frac{4\pi \sqrt{G}}{c^2}\sigma U^3.  \label{sithree}
\end{equation}
In the axially-symmetric case the equilibrium of charged dust was
investigated exhaustively by Bonnor \cite{fitwo} both in the Newtonian
theory and in general relativity. Eqs. (31) acquire terms with $\mu $ and $%
\sigma $ and serve as their expressions through $\phi $ and $f$. The
equilibrium condition reads 
\begin{equation}
\sqrt{G}\mu \nabla \sqrt{f}+\sigma \nabla \phi =0  \label{sifour}
\end{equation}
and leads directly to an arbitrary functional dependence $f^{1/2}=F\left(
\phi \right) $. Inserting it into the field equations yields the analogue of
Eq. (34) 
\begin{equation}
F\left( F_\phi ^2-1\right) \Delta \phi +\left( FF_{\phi \phi }-F_\phi
^2+1\right) F_\phi \nabla \phi \nabla \phi =0.  \label{sifive}
\end{equation}
When $\phi $ is found, $f$ is also known, while 
\begin{equation}
\frac{4\pi G}{c^2}\mu e^{2k}=FF_\phi \Delta \phi +\left( FF_{\phi \phi
}-F_\phi ^2-1\right) \nabla \phi \nabla \phi ,  \label{sisix}
\end{equation}
\begin{equation}
\frac{4\pi \sqrt{G}}{c^2}\sigma e^{2k}=-F\Delta \phi +2F_\phi \nabla \phi
\nabla \phi  \label{siseven}
\end{equation}
determine $\mu $ and $\sigma $. The analogues of Eqs. (32,38) become 
\begin{equation}
k_r=r\left( F_\phi ^2-1\right) F^{-2}\left( \phi _r^2-\phi _z^2\right)
,\quad k_z=2r\left( F_\phi ^2-1\right) F^{-2}\phi _r\phi _z.  \label{sieight}
\end{equation}

One solution of Eq. (65) is $F_\phi ^2=1$. It leads to the MP solutions with
dust, Eqs. (61-63). What is more interesting is that when $F_\phi ^2\neq 1$
one can define 
\begin{equation}
\Psi =\int \left( F_\phi ^2-1\right) ^{1/2}F^{-1}d\phi  \label{sinine}
\end{equation}
and Eq. (65) becomes $\Delta \Psi =0$, so $\Psi $ is analogous to $\psi $
but the function $F\left( \phi \right) $ is arbitrary. After it is fixed we
can find $\Psi $ and express everything in terms of it. Such solutions do
not satisfy Eq. (61) and are singular \cite{fithree}. Some of them have, in
addition, negative mass density, but in the remarkable example \cite{fitwo}
of pure root gravity $f=C_1\phi $ there is a region of positive $\mu $.

Charged dust clouds of spherical or spheroidal shape and obeying Eq. (63)
have been studied extensively in an astrophysical context \cite
{fifour,fifive,fisix,fiseven,fieight}. They have a number of interesting
properties when compared to usual stars: their mass and radius may be
arbitrary, very large redshifts are attainable, their exteriors can be made
arbitrarily near to the exterior of extreme charged black holes. In the
spherical case the average density can be arbitrarily large, while for any
given mass the surface area can be arbitrarily small. When their radius
shrinks to zero, many of their characteristics remain finite and non-trivial.

Some more general solutions of Eq. (63) have been given too \cite
{finine,sixty}. They include the case of constant $\mu $ with $U$ given in
terms of a Jacobi elliptic function. The idea that $\mu $ may be
concentrated on surfaces was also discussed \cite{sione}. Thin dust shells
of spherical, cylindrical or plane shape were given as examples. Finally, a
magnetostatic dust solution with functional dependence between $f$ and $%
A_\varphi $ is also known \cite{sitwo}.

The charge density required to satisfy Eq. (61) is quite small. It is
sufficient that in a sphere of neutral hydrogen one atom in about $10^{18}$
had lost its electron \cite{fifour}. However, all charged dust models have
one essential flaw: the equilibrium is very delicate and unstable and a
slight change in $\sigma $ would cause the cloud to expand or contract. On
the other hand, the models discussed in the present paper do not depend on
equilibrium conditions. They require just the creation of strong enough
electric or magnetic fields. Of course, if the charged cloud is divided into
parts with positive and negative particles (or ions), this can also induce
EM-fields and root gravity.

The appearance of the WMP relation (14) was studied also when the dust is
pressurized, i.e., in the case of charged perfect fluids \cite
{sithree,sifour} and a gravitational model for the electron was put forth 
\cite{sifive}. Like the case of charged dust, many different relations
between $f$ and $\phi $ are possible, but only partial results in the
spherical case were obtained \cite{sisix}. Spherical charged perfect fluid
interior solutions are reviewed in Ref. \cite{siseven}.

The analogy between electrovac and stationary fields is also worth being
mentioned. When there are no EM-fields, the equations following from the
metric (56) become \cite{sieight,sinine} 
\begin{equation}
2\Delta u=-e^{-4u}\nabla \Omega \nabla \Omega ,\quad \Delta \Omega =4\nabla
u\nabla \Omega ,  \label{seventy}
\end{equation}
\begin{equation}
R_{ab}^{\left( 3\right) }=2u_au_b+\frac 12e^{-4u}\Omega _a\Omega _b,
\label{seone}
\end{equation}
\begin{equation}
\Omega _a=\frac 12e^{4u}\sqrt{\gamma }\varepsilon _{abc}h^{bc},\quad
h_{ab}=\omega _{a,b}-\omega _{b,a}.  \label{setwo}
\end{equation}

The comparison between Eqs. (70-72) and Eqs. (57,58) yields the Bonnor
transformation \cite{three,foone,sinine} between electrovac and stationary
solutions 
\begin{equation}
f=f_s^2,\quad \phi =i\Omega ,\quad k=4k_s,  \label{sethree}
\end{equation}
given here for axially-symmetric fields. In the general case Eqs. (58,71)
coincide after scaling the coordinates, while preserving $\gamma _{ab}$ \cite
{sinine}. Eq. (72) simplifies for axial symmetry, 
\begin{equation}
\Omega _r=\frac{f_s^2}r\omega _z,\quad \Omega _z=-\frac{f_s^2}r\omega _r.
\label{sefour}
\end{equation}
For magnetovacs the transformation involves either $\lambda =i\Omega $ or
simply $\chi =i\omega $ which is a relation between the true potentials. One
obtains imaginary EM-fields which should be made real by a choice of the
integration constants.

When the Bonnor transformation is executed upon Weyl electrovacs, one finds
a differential system for the gravitomagnetic potential 
\begin{equation}
\omega _r=-ir\psi _z,\quad \omega _z=ir\psi _r.  \label{sefive}
\end{equation}
Its solution is $\omega =r\zeta _r$, $\zeta _z=i\psi $ with $\zeta $ being
another harmonic function. Hence, $\psi $ must be imaginary and the case $%
D<0 $ should be used. Eq. (41) transforms into 
\begin{equation}
f_s^{-1}=\cosh \frac{\sqrt{-D}}2\zeta _z+\frac b{\sqrt{-D}}\sinh \frac{\sqrt{%
-D}}2\zeta _z  \label{sesix}
\end{equation}
with $B=ib$, $-D=b^2+4$. This is exactly the Papapetrou (P) stationary
solution \cite{seventy,seone}, which has puzzled the researchers for ten
years (1953-1963) with its absent mass term, till the properly behaving Kerr
solution was found.

Stationary fields have been studied much more profoundly \cite{three}
because of their astrophysical importance. With the help of the Bonnor
transformation many facts about them and their Ernst potential $%
E_s=f_s+i\Omega $ hold also for electro(magneto)vacs. For example, one of
the degeneracies in $R_{ab}^{\left( 3\right) }$ leads to the functional
dependence $f_s\left( \omega \right) $ and the P-solution \cite
{setwo,sethree,sefour,sefive}, while another one leads to algebraically
special fields \cite{sesix}. The class of metrics with shearing geodesic
eigenrays and spin coefficient $\tau =0$ again gives the P-solution \cite
{seseven}. There are several theorems \cite{three}, Sec. 18.7, about the
subclass of conformastationary spacetimes \cite{seeight,senine,eighty} 
\begin{equation}
ds^2=f_s\left( dx^0+\omega _adx^a\right) ^2-\Lambda ^4\left( x,y,z\right)
\left( dx^2+dy^2+dz^2\right) .  \label{seseven}
\end{equation}
According to them, conformastats are of Petrov type D. All such metrics,
however, are explicitly known and are axially-symmetric. On the other side,
axisymmetric conformastats cannot have the so-called Ernst coordinates \cite
{eighty} $E_s$ and $E_s^{*}$ because $E_s=E_s\left( E_s^{*}\right) $. This
dependence is $f_s\left( \omega \right) $ in disguise, so all such solutions
belong to the P-class.

Finally, it is known that when $f_s$ and $\Omega $ are not functionally
dependent, Eq. (71) is more important than Eq. (70), which follows from it 
\cite{eione}. One can write a system of differential equations for $\gamma
_{ab}$ and expressions for $f,\Omega $ \cite{eitwo,eithree}, both for a
timelike or a spacelike Killing vector. In the axially-symmetric case an
equation of fourth differential order for $k$ results \cite{eifour}. When
applied to electrovacs, this shows that root gravity and usual gravity are
complementary. In non-Weyl fields $k$ (which is always proportional to $%
\kappa $) plays the role of a master potential, like $\psi $ (which is
proportional to $\sqrt{\kappa }$) in Weyl fields. One is tempted to
speculate that the purely electrically induced Weyl fields should bear no
reference to $k$ and the spacetime must be conformastatic ($B=2$).

\section{Electrostatic examples}

In this section four simple experimental set-ups are given where the effects
of root gravity may be detected and measured. They involve plane, spherical,
spheroidal and cylindrical symmetry. In these cases Weyl fields represent
the most general electrovac solution and there is no ambiguity as to their
appearance.

\subsection{The moving capacitor}

The study of plane-symmetric electrovac metrics goes back to 1926 \cite
{eifive}, but for a long time their Weyl nature remained unrecognized. A
plane-symmetric example of a WMP field was given first by Papapetrou \cite
{twelve}. Later Bonnor studied fields with $\phi =\phi \left( z\right) $,
which include also some non-Weyl solutions \cite{ninet}. Let us discuss the
gravitational field of a uniformly charged plane with charge density $\sigma 
$. The classic Maxwell potential (which becomes the master-potential) has
both plane and mirror symmetry 
\begin{equation}
\psi =-2\pi \sigma \left| z\right| \equiv q\left| z\right| .  \label{seeight}
\end{equation}
It vanishes at the plane $z=0$ and goes to infinity when $\left| z\right|
\rightarrow \infty $. In the case $D=0$ we should replace Eq. (78) into Eq.
(39). Kar has proposed a coordinate system where $\phi $ becomes harmonic, $%
\phi =q\left| z^{\prime }\right| $. It gives a constant electric field and
further deepens the analogy with classical electrostatics. Then the interval
becomes 
\begin{equation}
ds^2=f\left( dx^0\right) ^2-f^{-1}\left( dr^2+r^2d\varphi ^2\right)
-f^{-3}\left( dz^{\prime }\right) ^2,  \label{senine}
\end{equation}
where $f=\left( 1+q\left| z^{\prime }\right| \right) ^2$. When $D\neq 0$ one
sees from Eq. (38) that $k$ depends on $r$ and in fact 
\begin{equation}
k=-\frac D8q^2r^2.  \label{eighty}
\end{equation}
The metric does not inherit the symmetry of its source. Now the Kar's gauge
gives $\phi \left( z\right) =-B/2+q\left| z^{\prime }\right| +\alpha $.
Requiring flatness at $z^{\prime }=0$ \cite{ninet} one obtains $\alpha =B/2$
and Eqs. (40,41) yield 
\begin{equation}
ds^2=f\left( dx^0\right) ^2-e^{q^2\left( 1-\alpha ^2\right) r^2}\left(
f^{-1}dr^2+f^{-3}dz^{\prime 2}\right) -f^{-1}r^2d\varphi ^2,  \label{eione}
\end{equation}
where $f=1+2q\alpha \left| z^{\prime }\right| +q^2z^{\prime 2}$. The
non-inheritance is obvious again. The $D=0$ case is obtained when $\alpha
=\pm 1$ and is the only one with plane symmetry. This is our main argument
that $B=2$. Other values of $B$ (including $B=0$) introduce, in addition,
the parameter $\alpha $, which does not originate from the electric field,
unlike the charge parameter $q$. Probably it is invoked by mass sources. As
for the condition $f\left( z=0\right) =1$, its necessity will become clear
in the following. In conclusion, the only plane-symmetric Weyl metric is
given by Eq. (79). Another example of non-inheritance was given in Ref. \cite
{eisix}, where the metric is not rotationally invariant. The opposite
situation is also possible - plane-symmetric metric, associated with
semi-plane-symmetric EM-fields\cite{eiseven}.

In order to obtain a regular global solution, one must satisfy the junction
conditions \cite{eieight} at the plane. The metric is continuous there. The
extrinsic curvature reads 
\begin{equation}
K_{aa}=\frac 12\left| g_{zz}\right| ^{-1/2}\left( g_{aa}\right) _z,
\label{eitwo}
\end{equation}
where $a=x^0,\varphi ,r$ and there is no summation. Its eventual jump at the
plane determines the required energy-momentum tensor $S_{ab}$ of the massive
surface layer 
\begin{equation}
\kappa S_{\;b}^a=\gamma _{\;b}^a-\delta _{\;b}^a\gamma ,\quad \gamma
_{\;a}^a=K_{\;a}^a|_{-}^{+},\quad \gamma =\sum\limits_a\gamma _{\;a}^a.
\label{eithree}
\end{equation}
For the Weyl form of the axially-symmetric metric one has 
\begin{equation}
\kappa S_{\;0}^0=e^{u-k}\left( 2u_z-k_z\right) |_{-}^{+},\quad \kappa
S_{\;\varphi }^\varphi =-e^{u-k}k_z|_{-}^{+},\quad \kappa S_{\;r}^r=0.
\label{eifour}
\end{equation}

At first sight, the jump in the acceleration $u_z$ requires the introduction
of mass on the charged plane. However, its true cause is the electric field
present in the space around the plane and induced by the charge distribution
on it. This is seen after one writes the Einstein and Maxwell equations, Eq.
(31), in the present case 
\begin{equation}
u_{zz}=e^{-2u}\phi _z^2,\quad \phi _{zz}=2u_z\phi _z  \label{eifive}
\end{equation}
and makes use of Eqs. (14,37) 
\begin{equation}
u_z=\left( 1+\phi \right) \psi _z.  \label{eisix}
\end{equation}
Obviously, the jump in $u_z$ at $z=0$ is due to the jump in the
master-potential $\psi _z$ and consequently to the presence of charge. When
we integrate the first equation in Eq. (85) across the thickness of the
plane $\delta $, the l.h.s. has a jump, no matter how small $\delta $ is.
The r.h.s. will be an integral of $T_{\;0}^0$ which in the limit $\delta
\rightarrow 0$ coincides with $S_{\;0}^0$. Therefore, it also has a jump due
to the charge. The functional dependence $u\left( \phi \right) $
intermingles the equations in Eq. (85) and the charged surface layer of the
plane causes the jump due to $u$ in the extrinsic curvature too. For Weyl
fields the usual procedure of introducing mass and pressures on the surface
represents an attempt to model the influence of the electric field upon the
metric by traditional mass sources \cite{thfour,thfive,einine}. When $\psi $
from Eq. (78) is replaced in $g_i$, the plane will be attractive for one
sign of $q$, as if there was positive mass on it, but for the other sign of $%
q$ it will be repulsive as if it were made of exotic negative mass, which
breaks the energy conditions. There is no paradox because the true creator
of these effects is the energy-momentum tensor of the electric field, which
always satisfies all three energy conditions. As a result, the term $u_z$ in 
$S_{\;0}^0$ should be absent. In our case $k_z=0$ and even $k=0$, so there
is no mass surface layer at all.

Now let us put a second plane at $z=d$, charged in the opposite way. The
electric field is confined between the two planes and $\psi $ reads 
\begin{equation}
\psi =\frac{\psi _2}dz,\quad E_z=-f\bar \psi _z.  \label{eiseven}
\end{equation}
Here $\psi _2$ is the potential of the second plane ($\psi _1=0$). It is
related to the charge density by 
\begin{equation}
\psi _2=2\pi \sigma \frac d\varepsilon ,  \label{eieight}
\end{equation}
where $\varepsilon $ is the dielectric constant. Up to now we have
considered the case $\varepsilon =1$. More generally, $\varepsilon $ enters
Eq. (3) because the energy $T_{00}\sim \varepsilon E_i^2$ and consequently $%
T_{\mu \nu }\rightarrow \varepsilon T_{\mu \nu }$. When $\phi $ absorbs the
constants in Eq. (12) it will pick also $\sqrt{\varepsilon }$. The same is
true for $\psi $. The acceleration formula (49) becomes in a dielectric
medium 
\begin{equation}
g_z=\sqrt{G\varepsilon f}\frac{\bar \psi _2}d\approx 2.58\times 10^{-4}\frac{%
\sqrt{\varepsilon }}d\bar \psi _2.  \label{einine}
\end{equation}

Taking finite disks instead of planes one obtains the usual capacitor. The
field around its centre will be plane-symmetric while at the rim it will
depend on $r$ too. This may be diminished by careful electric shielding of
the capacitor. Outside there will be a vacuum flat metric which joins
smoothly the interior due to the conditions $f=const$ on the plates. One
comes to the conclusion that the capacitor will be subjected to practically
constant gravitational force $F_g$ in the $z$-direction, 
\begin{equation}
F_g=\sqrt{G\varepsilon }\frac Md\bar \psi _2=\sqrt{G\varepsilon }\mu S\bar 
\psi _2,  \label{ninety}
\end{equation}
where $M$ is the mass of the dielectric, $\mu $ is its mass density and $S$
is the area of the plate. This force is very different from the electric
force, trying to bring the plates together 
\begin{equation}
F_E=\frac{\varepsilon ^2S\bar \psi _2^2}{2\pi d^2}.  \label{nione}
\end{equation}
The latter is neutralized by the mechanical construction of the capacitor.
If it is hanging freely, the effect of $F_g$ may be tested experimentally.
To increase the acceleration it is advantageous to make $d$ small (typically 
$0.1cm\leq d\leq 1cm$), to raise the potential difference $\psi _2$ between
the plates up to $2\times 10^4CGS$ and to take a material with high $%
\varepsilon $. Some examples are gases ($\varepsilon \approx 1$), quartz ($%
4.5$), glycerine ($56.2$), water, muscles, electric ceramics ($81$), rutile (%
$TiO_2$) with $\varepsilon =170$. Ferroelectrics are even better if one can
cope with their hysteresis and tendency for saturation in strong fields.
Barium titanate ($BaTiO_3$), potassium dihydrogen phosphate($KH_2PO_4$) and
many others have $\varepsilon $ in the range of $10^4$. Thus $\sqrt{%
\varepsilon }/d$ may reach in principle $10^3$ and the maximum acceleration $%
g_{z,\max }=5.2g_e$ is more than enough to counter Earth's gravity. In a
more modest attempt one can take $\bar \psi _2=100kV$ and $\sqrt{\varepsilon 
}/d=10^2$ to get about one percent of $g_e$. Curiously, the same factors $%
\varepsilon ,d,\bar \psi _2$ are even more important in $F_E$, which is
always much bigger than $F_g$.

As we shall see, root gravity effect is strongest in this first capacitor
example and, more generally, when the electric (magnetic) field lines are
parallel. The gravitational acceleration is not strictly constant, having an
unobservable dependence on $z$. One cannot obtain root gravity terms by
studying fields with constant acceleration \cite{ninety}. Earth's field also
shares these ''shortcomings'' of artificial gravity: the acceleration's
directions are not parallel but meet at the planet's centre and its
magnitude decreases with height. In the traditional approach to the charged
plane its metric depends directly on $t$ and $z$, bypassing the
axially-symmetric step, see \cite{three}, p234. The Weyl nature is then
completely obscured. Some plane metrics are induced by null EM-fields, like
the special pp-wave given by Eq. (15.18) from \cite{three} or the
Robinson-Trautman solution, given by Eq. (28.43) from the same reference.
They are non-static because null EM-fields are incompatible with static
metrics, \cite{three}, Theorem 18.4, \cite{nione}. We also do not discuss
the non-null homogenous and uniquely conformally flat Bertotti-Robinson
solution, \cite{three} Sec.12.3. All other plane-symmetric solutions with
non-null EM fields have been found by Letelier and Tabensky \cite{nitwo}.
The metric is either static or spatially homogenous. We are interested in
the static branch. It has been reviewed in Ref. \cite{nithree}, however, the
electric field was not discussed and the ties with the Weyl fields remained
unelucidated. Let us clarify this issue now.

The metric reads in cylindrical coordinates 
\begin{equation}
ds^2=Kc^2dT^2-N\left( dR^2+R^2d\Phi ^2\right) -PdZ^2,  \label{nitwo}
\end{equation}
where $K,N,P$ depend on $Z$ and there is one relation between them. It
allows to express $K$ and $P$ as functions of $N$ 
\begin{equation}
K=\frac 1N\left( \beta _1+\beta _2\sqrt{N}\right) ,\quad P=\frac{N_z^2}{4\mu
^2}\left( \beta _1+\beta _2\sqrt{N}\right) ^{-1},  \label{nithree}
\end{equation}
\begin{equation}
\beta _1=\frac{\eta ^2}{4\mu ^2},\quad \beta _2=1-\beta _1,  \label{nifour}
\end{equation}
where $\eta $ and $\mu $ are constants related to the charge and mass
density on the plane. For the electric field we get from Eq. (5) 
\begin{equation}
E_z=-\phi _z=-\frac{\eta \sqrt{KP}}{2N},\quad \phi =\frac \eta {2\mu }\left( 
\frac 1{\sqrt{N}}-1\right) .  \label{nifive}
\end{equation}
We have chosen the boundary conditions $\phi \left( 0\right) =0$, $N\left(
0\right) =1$. It is easily seen that 
\begin{equation}
K=1+B_0\phi +\phi ^2,\quad N=\left( 1+\frac{2\mu }\eta \phi \right) ^{-2},
\label{nisix}
\end{equation}
\begin{equation}
B_0=\frac{2\mu }\eta +\frac \eta {2\mu }=\frac{2-\beta _2}{\sqrt{1-\beta _2}}%
.  \label{niseven}
\end{equation}
Thus $K$, the analogue of $f$, obeys Eq. (14) and root gravity terms are
present. When $\phi $ is turned off by $\eta \rightarrow 0$ ($\beta
_1\rightarrow 0$), $B_0$ does not stay constant. It increases to infinity
instead, so that $B_0\phi $ remains finite. This triggers the ''mass out of
charge'' mechanism described in Sec.3 and we end with the vacuum solution
for a massive plane with $K=N^{-1/2}$ \cite{nifour}. Hence, the mass is not
entirely of electromagnetic origin and the parameter $\mu $ is independent
from $\eta $ in general. Another limit is $\beta _2\rightarrow 0$. This was
explored first by McVittie \cite{nifive}. Then $\beta _1=1$, $\eta =\pm 2\mu 
$, $B_0=\pm 2$, $K=1/N$.

One can absorb $N$ in Eq. (92) by the change $Z\rightarrow N$. However, it
is better to determine $N$ from the relation between $K,P$ and $N$ which
specifies the coordinate system. The conformastatic spacetimes studied by
Weyl \cite{fourt}, Papapetrou \cite{twelve} and Bonnor \cite{ninet} have $%
N=P $. Kar \cite{eifive} used two gauges, $KP=1$ and $KP=N^2$ in his
pioneering work. The second gives constant $E_z$, as seen from Eq. (95).
McVittie \cite{nifive} worked in the gauge $KP=N$. Patnaick \cite{nisix}
attacked the problem with $K=P$, which is the Taub's gauge in the vacuum
case \cite{niseven}. The same gauge was utilized by Letelier and Tabensky 
\cite{nitwo}. Unfortunately, the equations cannot be integrated explicitly
in this gauge except for the McVittie limit.

In order to make comparison between the axially-symmetric approach and the
traditional plane-symmetric approach one should use the gauge $N=P$. This
differential equation for $N$ is easily solved and yields 
\begin{equation}
N=P=\left( 1-\mu Z+\frac{\beta _2\mu ^2}4Z^2\right) ^2,\quad K=\frac 1N%
\left( 1-\frac{\beta _2\mu }2Z\right) ^2,  \label{nieight}
\end{equation}
\begin{equation}
\phi _z=\frac \eta {2N}\left( 1-\frac{\beta _2\mu }2Z\right) ,\quad \phi =%
\frac \eta {2\mu }\left[ \left( 1-\mu Z+\frac{\beta _2\mu ^2}4Z^2\right)
^{-1/2}-1\right] .  \label{ninine}
\end{equation}
These expressions coincide with the $B=2$ case, Eqs. (39,78) only in the
McVittie limit where $\beta _2=0$ and consequently $q=\mu =\eta /2$, $\eta
=-4\pi \sigma $. In this limit the fields in Kar's gauge \cite
{eifive,nithree} coincide with Eq. (81) for $\alpha =1$ after the
identification $Z=z^{\prime }$. The original McVittie's metric is obtained
by passing to $e^{qz^{\prime \prime }}=qz^{\prime }+1$ 
\begin{equation}
ds^2=e^{2qz^{\prime \prime }}\left( dx^0\right) ^2-e^{-2qz^{\prime \prime
}}\left( dx^2+dy^2\right) -e^{-4qz^{\prime \prime }}\left( dz^{\prime \prime
}\right) ^2.  \label{hundred}
\end{equation}

In conclusion, the purely electric effect upon the metric of a charged plane
is given by the McVittie solution, while the deviation of $\beta _2$ from
zero ($\mu $ from $\eta /2$) signals the presence of mass in addition to the
charge. In this way one can increase (very inefficiently) $B_0$ and the root
gravity acceleration.

The solution in the general plane-symmetric case is also axially-symmetric.
There must exist a transformation of it to the Weyl interval, Eq. (20). The
latter is obtained from the general static element (see \cite{thirt}, Ch.8,
Sec.1) in isothermal form 
\begin{equation}
ds^2=g_{00}\left( dx^0\right) ^2+g_{\varphi \varphi }d\varphi
^2+g_{11}\left[ \left( dx^1\right) ^2+\left( dx^2\right) ^2\right] .
\label{hone}
\end{equation}
The interval in Eq. (92) is a particular case of Eq. (101) and, as expected, 
$\sqrt{KN}R$ satisfies the two-dimensional Laplace equation. Therefore, one
introduces the two-harmonic coordinate 
\begin{equation}
r\left( R,Z\right) =\sqrt{KN}R=R\left( 1-\frac{\beta _2\mu }2Z\right)
\label{htwo}
\end{equation}
and finds its conjugate $z\left( R,Z\right) $ from the condition $%
r+iz=F\left( R+iZ\right) $, $F$ being an analytic function. The
Cauchy-Riemann conditions lead to 
\begin{equation}
z=Z+\beta _3\left( R^2-Z^2\right) ,  \label{hthree}
\end{equation}
\begin{equation}
f=r^2R^{-2}\left( 1-\mu z+\beta _3R^2\right) ^{-2},\quad e^{2k}=r^2\left(
r^2+4\beta _3^2R^4\right) ^{-1},  \label{hfour}
\end{equation}
where $\beta _3=\beta _2\mu /4$ while $R^2$ is determined from the quadratic
algebraic equation 
\begin{equation}
4\beta _3^2R^4+\left( 1-4\beta _3z\right) R^2-r^2=0  \label{hfive}
\end{equation}
The addition of mass through the ''mass out of charge'' mechanism leads to $%
k\neq 0$, depending on both $r$ and $z$, while $f$ acquires $r$-dependence.
The plane $Z=0$ is transformed into the elliptic paraboloid of rotation $%
z=\beta _3r^2$. The non-electromagnetic mass changes the shape of the
charge-carrying surface when axially-symmetric coordinates are used.

\subsection{Repulsive gravity}

Let us study now the gravitational effect of electric fields with spherical
symmetry. The master potential is 
\begin{equation}
\psi =\frac q\rho ,\quad \rho ^2=r^2+z^2.  \label{hsix}
\end{equation}
This is the exterior solution for a metal conductive sphere of radius $\rho
_1$, charged to a potential $\psi _1=q/\rho _1$. Inside the sphere $\psi =0$
and the spacetime is flat. Outside we obtain the charged \cite
{nieight,ninine} Curzon \cite{hundred} solution. Eq. (38) gives 
\begin{equation}
k=-\frac{Dq^2r^2}{8\rho ^4}.  \label{hseven}
\end{equation}
Like in the plane-symmetric case the metric does not inherit the spherical
symmetry of $\psi $ unless $D=0$ ($B=2$). This corresponds to the critically
charged Curzon metric. Eq. (39) yields 
\begin{equation}
\phi =\frac q{\rho -q},\quad f=\left( 1-\frac{2q}\rho +\frac{q^2}{\rho ^2}%
\right) ^{-1}.  \label{height}
\end{equation}
The charge in CGS units $\bar q$ is connected to $q$ by $q=\frac{\sqrt{G}}{%
c^2}\bar q$. Eq. (49) gives for the acceleration 
\begin{equation}
g_\rho =-\sqrt{Gf}\frac{\bar \psi _1\rho _1}{\rho ^2}.  \label{hnine}
\end{equation}
It has a maximum at the sphere and utilizing our maximum potential we obtain 
$\left| g_{\rho ,\max }\right| =5.06/\rho _1$. When $\rho _1=10cm$, one gets 
$0.5cm/s^2$. Formula (82) for the extrinsic curvature holds in the present
case after the replacement $z\rightarrow \rho $ and $a=x^0$, $\theta $, $%
\varphi $, i.e., the axially-symmetric element is written in spherical
coordinates 
\begin{equation}
ds^2=e^{2u}\left( dx^0\right) ^2-e^{-2u}\left[ e^{2k}\left( d\rho ^2+\rho
^2d\theta ^2\right) +\rho ^2\sin ^2\theta d\varphi ^2\right] .  \label{hten}
\end{equation}
Expression (84) for $S_{\;0\text{ }}^0$ and $S_{\;\varphi }^\varphi $ holds
after the same change, while $S_{\;\theta }^\theta =0$. Once again we argue
that the $u_\rho $ term is in fact absent, while $k=0$ and there is no mass
surface layer, but only a charged one. The acceleration can be enhanced by
introducing a second sphere with $\rho _2>\rho _1$ and filling the space
between the two spheres with high $\varepsilon $ dielectric. A spherical
condenser is obtained and the master potential is increased by $\sqrt{%
\varepsilon }\rho _2\left( \rho _2-\rho _1\right) ^{-1}$. As in the previous
case, for one sign of $q$ gravity becomes repulsive. The interaction between
massive bodies with repulsive (negative mass) and attractive (positive mass)
gravitation has been discussed in Refs \cite{hone,htwo}. In our case
repulsion is a natural property of the electric field and does not break the
energy conditions.

Let us deform the charged sphere into an oblate or prolate spheroid. Its
gravitational field is described best in the corresponding spheroidal
coordinates $x,y$ (to be distinguished from the cartesian coordinates in the
previous sections) 
\begin{equation}
r=\tau \left( x^2\pm 1\right) ^{1/2}\left( 1-y^2\right) ^{1/2},\quad z=\tau
xy,  \label{heleven}
\end{equation}
where $\tau $ is a parameter. In the prolate case one has 
\begin{equation}
x=\frac 1{2\tau }\left( l_{+}+l_{-}\right) \equiv \frac L\tau ,\quad y=\frac 
1{2\tau }\left( l_{+}-l_{-}\right) ,  \label{htwelve}
\end{equation}
\begin{equation}
l_{\pm }=\sqrt{\left( z\pm \tau \right) ^2+r^2}.  \label{hthirt}
\end{equation}
The master potential is taken to depend only on $x$ ($x=x_1$ is the surface
of the charged spheroid). The harmonic solution is given by the Legendre
function $Q_0\left( x\right) $ 
\begin{equation}
\psi =\frac q2\ln \frac{x-1}{x+1}  \label{hfourt}
\end{equation}
and generates the general solution of the Einstein-Maxwell equations for
such symmetry. From Eq. (48) one has $q=4\pi \sigma \left( x_1^2-1\right) $.
This potential coincides in form with the Weyl rod for the vacuum $\gamma $%
-metric \cite{twthree}. Eq. (38) gives 
\begin{equation}
k=\frac{Dq^2}8\ln \frac{x^2-1}{x^2-y^2},  \label{hfift}
\end{equation}
which depends on $y$ too, signalling non-inheritance, except when $B=2$. The
same conclusion follows in the oblate case, but there $\psi =-q\arctan 1/x$.
Repulsive gravity also arises in these cases.

It is well-known that the unique spherically symmetric electrovac solution
with mass $M$ and charge $\bar Q$ is the Reissner-Nordstr\"om solution \cite
{hthree,hfour} 
\begin{equation}
ds^2=\left( 1-\frac{2m}R+\frac{Q^2}{R^2}\right) c^2dT^2-\left( 1-\frac{2m}R+%
\frac{Q^2}{R^2}\right) ^{-1}dR^2-R^2\left( d\Theta ^2+\sin ^2\Theta d\Phi
^2\right) ,  \label{hsixt}
\end{equation}
\begin{equation}
\phi =\frac QR,\quad m=\frac{GM}{c^2},\quad Q^2=\frac{G\bar Q^2}{c^4}.
\label{hsevent}
\end{equation}
Hence, at least the solution given by Eq. (108) should be transformable into
the RN solution. Surprisingly, all three solutions described above are
formally equivalent to its three cases; undercharged ($Q^2<m^2$), critically
(extremely) charged ($Q^2=m^2$) and overcharged ($Q^2>m^2$). The proof uses
the Weyl form of the RN solution \cite{foseven,ninine}. The essential step
is to apply Kar's gauge, in which $\phi $ becomes harmonic instead of $\psi $%
. When $D=0$ we transform Eq. (108) by setting $R=\rho -q$, $\cos \Theta
=z/\rho $ and obtain Eqs. (116,117) with $m=-q$, $Q=q$. The sphere $\rho
=\rho _1$ transforms into the sphere $R=R_1=\rho _1-q$. When $D>0$ we take
Eq. (114) and fix $q$ by the condition $Dq^2/4=1$. Then we utilize Eqs.
(42,43) to find $\phi $ and $f$ 
\begin{equation}
\phi =-\frac 2{\sqrt{D}\left( \frac B{\sqrt{D}}+x\right) },\quad f=\frac{%
x^2-1}{\left( \frac B{\sqrt{D}}+x\right) ^2}.  \label{heightt}
\end{equation}
Making the identifications 
\begin{equation}
\tau =\left( m^2-Q^2\right) ^{1/2},\quad B=-\frac{2m}Q,  \label{hninet}
\end{equation}
we obtain (together with Eqs. (111-115)) the formulas for the undercharged
case from Ref. \cite{foseven}. The transformation 
\begin{equation}
R=\tau x+m,\quad \cos \Theta =y  \label{htwenty}
\end{equation}
maps this solution into the RN solution. The charged spheroid $x=x_1$ goes
into the charged sphere $R=R_1=\tau x_1+m$. It is seen from Eq. (119) that $%
B\rightarrow \infty $ when $Q\rightarrow 0$ and $B\phi $ remains finite. The
independent mass parameter $m$ again arises from the ''mass out of charge''
mechanism. A similar chain of arguments connects the oblate spheroid case to
the overcharged RN solution.

In conclusion, the electrically induced spherically symmetric gravitational
field is given by the critically charged Curzon solution, which is
equivalent to an extreme RN solution with $m=-Q$. There is a transformation
of RN solutions with $m\neq \pm Q$ into spheroidal metrics with particular $%
D $. The distance between the spheroid's foci $\tau $ is related to the
deviation of $m$ from its electromagnetic value. Thus part of the exterior
solution's mass is from electromagnetic origin. This has been long known for
interior charged perfect fluid solutions \cite{siseven}. The results are
analogous to the plane-symmetric ones. The RN solution also exerts a
repulsive force for certain values of its parameters \cite{hfive,hsix}. It
appears, however, due to another mechanism. In the charged Curzon solution
the sign of the charge is decisive and the region with repulsion occupies
the whole exterior space. In the RN solution the mass is always positive,
but enters the metric with a negative sign, so that a competition with the
charge term is possible. The region with repulsion is finite, $R<Q^2/m$.

\subsection{The microgravity chamber}

It is natural to ask whether artificial gravity of electromagnetic origin
can be created in a big enough chamber for people to work inside. This can
be achieved by taking a capacitor and increasing the distance $d$ between
the plates to, let us say, $2m$. The dielectric should be some gas, $%
\varepsilon \approx 1$ and we cannot use its enhancing property. Since air
breaks at $E_{crit}=3\times 10^4V/cm=100CGS$, we better use vacuum. Taking $%
\bar \psi _{\max }$ one creates microgravity, $g_z=2.58\times 10^{-2}cm/s^2$%
. One can compete with $g_e$ when $\bar \psi _2=3.8\times 10^4\bar \psi
_{\max }$. In addition, the exact solution is very complicated. Although the
effect is tiny, it is worth to present a simple exact solution with the same
properties.

The general solution of the Laplace equation in spherical coordinates can be
written, separating the variables, as 
\begin{equation}
\psi =\sum_{n=0}^\infty \left( C_{1n}\rho ^n+C_{2n}\rho ^{-\left( n+1\right)
}\right) P_n\left( \cos \theta \right) .  \label{htwone}
\end{equation}
Let us take a charged sphere of radius $\rho _1$. When $n=1$ the interior
solution has almost constant $E_z$%
\begin{equation}
\psi ^{-}=q\rho \cos \theta .  \label{htwtwo}
\end{equation}
The potential will be continuous when $n=1$ also for the exterior solution 
\begin{equation}
\psi ^{+}=\frac{q\rho _1^3\cos \theta }{\rho ^2},\quad \sigma =\frac{3q}{%
4\pi }\cos \theta .  \label{htwthree}
\end{equation}
Here $\sigma $ is given by the jump of $\psi _\rho $ at the sphere.
Obviously $\psi ^{+}$ is a dipole solution. In classical electrostatics this
charge distribution arises in a number of cases:

1) Two oppositely charged balls with slightly displaced centres.

2) Uniformly polarized ball.

3) Metal uncharged ball in a uniform constant electric field.

4) Electric dipole in a spherical shell cut inside a conductor.

These set-ups are inapplicable for our purposes. One must create the cosine
law density by using plastic or other non-conductive material, since the
sphere is not an equipotential surface. Inside it is the plane-symmetric
solution with $B=2$ and $k=0$. Outside $k$ should vanish too, in order to be
continuous. The gravitational accelerations read 
\begin{equation}
g_z^{-}=\sqrt{G}q,\quad g_r^{-}=0,\quad g_z^{+}=\frac{\sqrt{G}q\rho
_1^3\left( r^2-2z^2\right) }{\rho ^5},\quad g_r^{+}=-\frac{3\sqrt{G}q\rho
_1^3rz}{\rho ^5}.  \label{htwfour}
\end{equation}
Outside the sphere there is a peculiar gravitational force $\sim \rho ^{-3}$%
. The monopole term is absent and one may think that the source is a dipole
of positive and negative mass particles. In fact, the source is the charge
of the sphere and, as in the capacitor example, gravitation is concentrated
mainly inside and rapidly decreases with the outside distance. Intuitively,
this is a more economic way than to generate long-distance monopole terms.

The traditional junction conditions yield 
\begin{equation}
\kappa S_{\;0}^0=2e^uu_\rho |_{-}^{+}=2e^{2u}\psi _\rho |_{-}^{+},\quad
S_{\;\theta }^\theta =S_{\;\varphi }^\varphi =0.  \label{htwfive}
\end{equation}
Again, the introduction of mass on the surface is not necessary - it will
only duplicate the effect of the charge.

Gravitational solutions corresponding to electric or magnetic dipoles plus
eventual mass distributions have been considered in the literature \cite
{five,hseven,height}. The global solution described here was discussed too 
\cite{hnine}. However, the spherical layer was made from dust, which
requires for equilibrium the introduction of mass according to the law (54).
Several definitions of acceleration were discussed in relativistic units,
where the root gravity effect is obscured. The focus was put on the
similarity between the interior solution and ''gravity'' in an accelerated
coordinate system, and the radiation of test-particles.

\subsection{Charged cylinder}

Finally, we discuss the gravitational field of an infinitely long charged
cylindrical shell of radius $r_1$. In the exterior 
\begin{equation}
\psi =q\ln \frac r{r_0},\quad \psi _r=\frac qr,  \label{htwsix}
\end{equation}
where $q=-4\pi \sigma r_1$ is determined by the charge per unit length $%
\sigma $ and $r_0$ is a scale not set by $\sigma $. Inside the cylinder $%
\psi =0$. The second gravitational potential reads 
\begin{equation}
k_r=\frac{Dq^2}{4r},\quad e^{2k}=r^{\frac{Dq^2}2}.  \label{htwseven}
\end{equation}
The symmetry of $\psi $ is inherited because $f,k$ depend only on $r$. The
extrinsic curvature is obtained from Eq. (82) by making the change $%
z\rightarrow r$. The same is true for $S_{\;a}^a$ and Eq. (84). Eq. (31)
becomes 
\begin{equation}
u_{rr}+\frac 1ru_r=e^{-2u}\phi _r^2,\quad \phi _{rr}+\frac 1r\phi
_r=2u_r\phi _r.  \label{htweight}
\end{equation}
Making use of Eqs. (14,37) we find an equation similar to Eq. (86), which
shows that $u_r$ has a jump induced by $\psi _r$. $T_{\;0}^0$ also acquires
a jump when Eq. (128) is integrated across the layer and it is exactly the $%
u_r$ term in $S_{\;0}^0$. Hence, this term is redundant and the surface
layer is given in terms of $k_r$%
\begin{equation}
\kappa S_{\;0}^0=\kappa S_{\;\varphi }^\varphi =-e^{u-k}k_r|_{-}^{+},\quad
\kappa S_{\;r}^r=0.  \label{htwnine}
\end{equation}
Put differently, a mass surface layer of this type (with equal mass and
tangential pressure) generates a non-zero $k$ in the metric, $D<0$ for
positive mass.

Studying purely electromagnetic effects on the metric, we set $D=0$. We also
demand that when the electric field is turned off ($q=0$), we should have $%
\phi =0$ and $f=1$. Therefore, in Eq. (39) $\psi _0=-1$ and 
\begin{equation}
\phi =-1+\left( 1-q\ln \frac r{r_0}\right) ^{-1},\quad f=\left( 1-q\ln \frac 
r{r_0}\right) ^{-2}.  \label{hthirty}
\end{equation}
The interval reads 
\begin{equation}
ds^2=\left( 1-q\ln \frac r{r_0}\right) ^{-2}c^2dt^2-\left( 1-q\ln \frac r{r_0%
}\right) ^2\left( dz^2+dr^2+r^2d\varphi ^2\right) .  \label{hthone}
\end{equation}
Comparing it to Eq. (22.16) from Ref. \cite{three}, one sees that the latter
refers to Eq. (22.4b) with $A=0$, instead of Eq. (22.4a). Thus, $t$ and $z$
should be interchanged. Second, the $a$ in $A_i$ (corresponding to our $q$)
should be in the denominator. Third, in Ref. \cite{three} the condition $%
\psi _0=0$ was used and the gauge transformation $\phi ^{\prime }=\phi +1$
was made. In this case $f=\phi ^{\prime 2}$ and there is no root gravity
term. However, when $q\rightarrow 0$ both the metric and $\phi $ become
ill-defined. Finally, Ref. \cite{hten} is quoted for this result. In fact,
Raychaudhuri applied the Rainich formalism to rederive some results of
Bonnor \cite{ninet} (who used the Weyl formalism) and whose Eq. (2.30) he
quoted.

In reality the charged cylinder is also massive, so that one must allow for
a tiny $k$ to appear. Eqs. (127,129) show that a mass density $m_c$ induces
negative $D$%
\begin{equation}
D=-\frac{2Gm_c}{\pi c^2\sigma ^2r_1}.  \label{hthtwo}
\end{equation}
It confirms our assertion that $D$ is generated by ''parasitic'' mass
sources. They can be neglected since $k_r\sim \kappa $ and the mass induced
acceleration for realistic $m_c$ is much smaller than the
electromagnetically induced one.

Let us discuss further the cases with $D\neq 0$ in order to make connections
with the existing literature. We introduce the notation 
\begin{equation}
n=\frac 12\sqrt{\left| D\right| }q,\quad \varsigma =n\ln \frac r{r_0},\quad
\Gamma _{\pm }=\frac 1{\sqrt{\left| D\right| }}\left[ \left( \frac r{r_0}%
\right) ^n\pm \left( \frac r{r_0}\right) ^{-n}\right] .  \label{hththree}
\end{equation}

When $D>0$ we use Eqs. (42-44) with $\psi _0=0$ to find 
\begin{equation}
\phi =-\frac B2-\frac{\sqrt{D}}2\frac{\Gamma _{+}}{\Gamma _{-}},\quad f=%
\frac D{4\sinh ^2\varsigma }=\Gamma _{-}^{-2},\quad e^{2k}=r^{2n^2}.
\label{hthfour}
\end{equation}
Using Eqs. (14, 37) one can express $\phi $ through $f$ and $\psi $ to
obtain in the present case 
\begin{equation}
\phi +\frac B2=\frac{f_i}{2f\psi _i}=-\frac{r\left( \Gamma _{-}\right) _r}{%
q\Gamma _{-}}.  \label{hthfive}
\end{equation}

When $D<0$, Eqs. (40-41) with $\psi _0=0$ give 
\begin{equation}
\phi =-\frac B2+\frac 12\sqrt{-D}\tan \varsigma =-\frac B2+\frac{\sqrt{-D}}2%
\frac{\Gamma _{-}\left( in\right) }{\Gamma _{+}\left( in\right) },
\label{hthsix}
\end{equation}
\begin{equation}
f=-\frac D{4\cos ^2\varsigma }=\left[ \Gamma _{+}\left( in\right) \right]
^{-2},\quad e^{2k}=r^{-2n^2}.  \label{hthseven}
\end{equation}
These trigonometric expressions were given by Bonnor \cite{ninet} and quoted
by Raychaudhuri \cite{hten}. We have explained already that the $D>0$ case
does not simply follow from the substitution $in\rightarrow n$, as stated in
Ref. \cite{ninet} and Eq. (134) is an illustration of this. Eqs. (134,135)
appear as Eq. (22.14) in Ref. \cite{three} where

\begin{equation}
a_1=\frac{r_0^{-n}}{\sqrt{D}},\quad a_2=-\frac{r_0^n}{\sqrt{D}},\quad
A_0=\phi +\frac B2  \label{htheight}
\end{equation}
and $b=-q$, $m=n$. The credit is given again to Ref. \cite{hten}. The same
formula was rederived in a recent paper \cite{heleven}. The particular case $%
n=1/2$ was studied in the pioneering work on the subject \cite{htwelve} (see
also Ref. \cite{hthirt} where the Rainich formalism was applied). In Ref. 
\cite{three} $n$ is considered real and the Bonnor's case $D<0$ is not
mentioned. Yet, when a Bonnor transformation is made, it maps this case on
the Lewis class of stationary cylindrical vacuum solutions, while the real $%
n $ case ($D>0$) is mapped onto the Weyl class (see Eq. (22.6) from Ref. 
\cite{three}). These two classes are preserved also in the electrovac case.
In addition, the metric (131) parallels the limiting Van Stockum class \cite
{seone,hfourt}. Like the $D=0$ case, the metrics (134,137) are singular
because $\psi _0=0$. They become regular when $\psi _0$ is chosen according
to Eqs. (40,44). The parameter $r_0$ in Eq. (131) can be set to $r_0=r_1$,
making the metric continuous at the shell. Like in the previous examples,
the cylinder's gravity is repulsive for one sign of $\sigma $.

It is easy to see that at $r=0$ the above solutions are singular, no matter
how one chooses $r_0$ and $D$. This is the reason to use them only as
exterior solutions. One can invoke the ''mass out of charge'' mechanism to
end with the vacuum cylindrical Levi-Civita (LC) solution \cite
{seven,twthree} 
\begin{equation}
ds^2=r^{4\nu }c^2dt^2-r^{8\nu ^2-4\nu }\left( dr^2+dz^2\right) -r^{2-4\nu
}d\varphi ^2.  \label{hthnine}
\end{equation}
For this purpose we set $\sqrt{D}=r_0^n$ and send $B$ and $D>0$ to infinity,
keeping $n$ finite. In this way one of the terms in $\Gamma _{\pm }$
vanishes and Eq. (139) results with $n=2\nu $. On the other side, finite $n$
means $q\rightarrow 0$, i.e. the electric field vanishes. $B$ does not stay
constant, consequently the term $B\phi $ is preserved and from Eqs. (7,135)
we get 
\begin{equation}
\frac{g_r}{c^2}=u_r=\left( \frac B2+\phi \right) \psi _r=-\frac{\sqrt{D}q}{2r%
}\frac{\Gamma _{+}}{\Gamma _{-}}\rightarrow \frac{2\nu }r.  \label{hforty}
\end{equation}
The LC metric describes the gravitational field of an infinite line mass
(ILM) and is singular at the axis. Eq. (140) shows that we have reproduced
its acceleration out of a charged massless solution.

\section{Magnetostatic examples}

We have pointed out that magnetic fields produce the same effects as
electric ones. One must replace $\phi $ by $\lambda $, $E_i$ by $H_i$, while 
$\psi $ becomes the magnetic scalar potential, determined by surface
currents. This is confirmed also by the definition of the magnetic field 
\cite{hfift} 
\begin{equation}
H_i=-\frac 12\sqrt{-g}\varepsilon _{ikl}F^{kl},  \label{hfoone}
\end{equation}
which gives, using Eq. (22) 
\begin{equation}
H_i=-\bar \lambda _i=-f\bar \psi _i.  \label{hfotwo}
\end{equation}
Eqs. (21,37) provide the connection between the scalar potential and the
only component ($\bar \chi =A_\varphi $) of the vector potential 
\begin{equation}
\bar \chi _z=r\bar \psi _r,\quad \bar \chi _r=-r\bar \psi _z.
\label{hfothree}
\end{equation}
It should be noted that in flat spacetime magnetostatics the formula $\vec H%
=curl\vec a$ involves the physical component in curvilinear coordinates $%
a_{\left( \varphi \right) }=A_\varphi /r$. In magnetogravity one has to find
the scalar potential anyway, in order to obtain $f$.

In a general medium the energy is $T_{00}\sim \mu H^2$, where $\mu $ is the
magnetic constant and consequently $T_{\mu \nu }\rightarrow \mu T_{\mu \nu }$%
. Hence, $\psi $ picks also the multiplier $\sqrt{\mu }$. Eq. (142) remains
unchanged. Arguments, analogous to those for electric fields, lead to the
conclusion that $B=2$ for a pure magnetic effect upon gravity. Formula (49)
for the acceleration becomes with high degree of precision 
\begin{equation}
g_i=-\sqrt{G\mu }H_i.  \label{hfofour}
\end{equation}
It is similar to the electric case, Eqs. (53,89). The gravitational force is
very different from the Lorentz force, acting upon charged particles 
\begin{equation}
\vec F_L=e\vec E+\frac{e\mu }c\left( \vec v\times \vec H\right) .
\label{hfofive}
\end{equation}
In it the electric force does not involve $\varepsilon $, while the magnetic
does involve $\mu $, but acts only on moving charges. There are terms,
depending on the velocity, also in $g_i$. We are interested on its effect
upon macroscopic bodies for which $v\ll c$ so we have neglected them and
study only acceleration at rest.

\subsection{Current loop}

The first magnetovac Weyl solution was given by Papapetrou \cite{twelve}.
Later Bonnor presented a number of examples \cite{five,twone}, including the
gravitational field of a current loop. He found a very small effect, based
on mass-energy considerations. The root gravity term remained unnoticed.

The magnetic field of a current loop is well-known and includes elliptic
integrals. On the $z$-axis and at the centre the field simplifies 
\begin{equation}
H_z\left( z,0\right) =0.2\pi I\frac{r_1^2}{\left( r_1^2+z^2\right) ^{3/2}}%
,\quad H_0\equiv H_z\left( 0,0\right) =\frac{0.2\pi I}{r_1},  \label{hfosix}
\end{equation}
where $r_1$ is the loop's radius in $cm$, $H$ is measured in Gauss and the
current $I$ is measured in $Amps$. Setting $\mu =1$ we get from Eq. (144) 
\begin{equation}
g_0\equiv \left| g_z\left( 0,0\right) \right| =1.62\times 10^{-4}\frac I{r_1}%
.  \label{hfoseven}
\end{equation}
Earth's acceleration is reached when $H_e=3.8\times 10^6G=380T$ or $%
I_e/r_1=6.05\times 10^6A/cm$. For a laboratory set-up let us take $r_1=100cm$%
. If a lightning with cross-section of radius $r_0=10cm$ circles around the
loop, one may take $I=10^5A$, the current density being $J=3.18\times
10^2A/cm^2$ and $g_0=0.162cm/s^2$. One can reach $g_e$ with a current of $%
6.05\times 10^8A$.

It is advantageous to make the loop thicker, turning it into a finite
solenoid. Let the inner radius be $r_1$, the outer radius be $r_2$ and the
height be $l$. Then \cite{hsixt,hsevent} 
\begin{equation}
H_0=F\left( \alpha ,\beta \right) r_1J,\quad F\left( \alpha ,\beta \right)
=0.4\pi \beta \ln \frac{\alpha +\sqrt{\alpha ^2+\beta ^2}}{1+\sqrt{1+\beta ^2%
}},  \label{hfoeight}
\end{equation}
where $\alpha =r_2/r_1$, $\beta =l/r_1$ and $J$ is the current density. As
an example, let us take $r_1=100cm$, $r_2=2r_1$, $l=2r_1$. Then $F\approx 1$
and $J=H_0/100$. Now $g_e$ is reached when $J_e=3.8\times 10^4A/cm^2$.

One can increase the acceleration by creating magnetic fields in a
ferromagnetic medium. Iron has $\mu _{\max }=5000$ ($\sqrt{\mu _{\max }}%
=70.7 $). There are alloys, like supermalloy (79\% Ni, 16\% Fe, 5\% Mo),
which have $\mu _{\max }=8\times 10^5$, $\sqrt{\mu _{\max }}=894.4$. Their
saturation field is comparatively low, $H_s=8\times 10^3G$ and the maximum
is obtained roughly for one third of this value. The effective field in Eq.
(144) will be $H_{eff}=\sqrt{\mu _{\max }}H_{\max }\approx 238T$, which is
of the order of $H_e$. Making a disc from this material with a current
flowing in a strip around the rim, we get the magnetic analogue of the
moving capacitor example from Sec.5.1.

There are two conceptual issues which need clarification. The metric depends
directly on the scalar potential, but $\psi $ is multivalued in
magnetostatics. If there is a current-carrying surface, the jump takes place
there and probably the metric can be made continuous in this region of
non-vanishing $J$. In the case of a current loop the jump can be arranged to
take place on any surface, based on the loop. Symmetry considerations
require to take the disk $z=0$, $\rho <r_1$ as such surface. In spherical
coordinates \cite{twone} 
\begin{equation}
\psi =0.2\pi J\left[ \pm 1-\frac \rho {r_1}P_1+\frac{\rho ^3}{2r_1^3}%
P_3-...+\left( -1\right) ^{n+1}\frac{1.3...\left( 2n-1\right) }{2.4...2n}%
\left( \frac \rho {r_1}\right) ^{2n+1}P_{2n+1}+...\right] ,  \label{hfonine}
\end{equation}
where the sign coincides with $signz$ and $P_n\left( \cos \theta \right) $
are the Legendre polynomials. Now, let us remember that we have been unable
to determine the sign of $B$ and for definiteness worked with positive $B$.
In fact, Eq. (39) should read 
\begin{equation}
f=\left( 1\pm \psi \right) ^{-2}.  \label{hfifty}
\end{equation}
Using different signs for $z$ positive or negative, and taking into account
that $P_{2n+1}\left( \cos \frac \pi 2\right) =0$, makes $f$ continuous at $%
z=0$. The derivatives of $\psi $ are single-valued, hence, $k$ is also
continuous. The acceleration will have a jump and change of sign at the disk
within the loop, like it has on both sides of a charged plane (disk). We
took the absolute value of $z$ in Eq.(78), which plays the same role. This
jump has been attributed to some mass surface layer \cite{twone}, which in
our view is not correct.

The second issue concerns the fact that the magnetic field far away from the
loop has a dipole character and the acceleration does not contain a monopole
term. As we have argued, this is not a tragedy, because electromagnetic
fields can create artificial gravity in a confined region of space.
Intuitively, this is more favourable energetically. Under a Bonnor
transformation such fields go into the Papapetrou stationary solution, which
possesses the same feature. Of course, the material of the loop always has
some mass, which induces a monopole term, negligible with respect to the
strong effect from root gravity.

\subsection{Spherical solenoid}

The spherical solenoid is an example of a closed current-carrying surface
and a magnetic analogue of the microgravity chamber discussed in Sec.5.3.
The scalar potential is similar to Eqs. (122,123) 
\begin{equation}
\bar \psi ^{-}=-H_0\rho \cos \theta ,\quad \bar \psi ^{+}=\frac{H_0\rho _1^3%
}{2\rho ^2}\cos \theta .  \label{hfione}
\end{equation}
$H_0$ is the approximately constant magnetic field inside the solenoid,
directed along the $z$-axis. The normal component $\bar \psi _\rho $ is
continuous at $\rho =\rho _1$ but $\bar \psi _\theta $ has a jump according
to Eq. (55) 
\begin{equation}
H_\theta |_{-}^{+}=-\frac 1{\rho _1}\bar \psi _\theta |_{-}^{+}=\frac{4\pi }c%
J.  \label{hfitwo}
\end{equation}
Eq. (151) gives 
\begin{equation}
J=J_0\sin \theta ,\quad H_0=-\frac{0.8\pi }3J_0,  \label{hfithree}
\end{equation}
where $J_0$ is in $A/cm$. This kind of current can be obtained by rotating a
uniformly charged metal sphere or making a spherical solenoid with constant
number of coils per unit length of the $z$-axis. The last condition ensures
uniform magnetic field also inside a prolate or oblate spheroid. It can be
fulfilled much easier than the cosine law for $\sigma $ in Eq. (123).
Putting Eq. (153) into Eq. (144) yields 
\begin{equation}
g_z=2.17\times 10^{-4}J_0.  \label{hfifour}
\end{equation}
Earth's acceleration is reached when $J_0=4.51\times 10^6A/cm$. This is
better than a loop and worse than a fat solenoid because the sphere is thin,
but the gravitational field is uniform and mimics the field of Earth.
Outside the sphere the fields have dipole character and quickly vanish. The
junction conditions again lead to Eq. (125) but now $S_{\;0}^0\equiv 0$. No
mass surface layer is necessary.

This example was considered in Ref. \cite{eight}. There the physical
component $H_{\left( \theta \right) }$ was taken in Eq. (152), which is
wrong, although the difference is negligible. The current determines the
jump in the classical scalar potential, which is related to the classical
magnetic field $H_\theta $. Perturbation theory was used in this reference,
which starts with terms $\sim \kappa $ and misses the root gravity term $%
\sim \sqrt{\kappa }$ in the exact solution. It is no wonder that the results
of Ref. \cite{eight} confirm the findings for a long solenoid \cite{four},
given here in Eq. (9).

\subsection{Long solenoid}

In order to explain why Eq. (9) is so different from Eq.(17), some
preparatory work is necessary. Let us rewrite Eq. (31) in terms of $%
f,\lambda $%
\begin{equation}
\Delta \ln f=2f^{-1}\left( \lambda _r^2+\lambda _z^2\right) ,\quad \Delta
\lambda =f^{-1}\left( f_r\lambda _r+f_z\lambda _z\right) .  \label{hfifive}
\end{equation}
Next, let us introduce $h=r^2/f$ and express $\lambda $ through $\chi $. The
second equation is satisfied trivially. The first one and the compatibility
condition $\lambda _{rz}=\lambda _{zr}$ give 
\begin{equation}
\Delta \ln h=-2h^{-1}\left( \chi _r^2+\chi _z^2\right) ,\quad \Delta \chi
=h^{-1}\left( h_r\chi _r+h_z\chi _z\right) .  \label{hfisix}
\end{equation}
These equations differ from Eq. (155) only by a minus sign. Let us search
for Weyl solutions in the system $h,\chi $ by demanding $h=h\left( \chi
\right) $. The analogues of Eqs. (14,34) are 
\begin{equation}
f=\frac{r^2}{C_1+B_1\chi -\chi ^2},\quad \Delta \chi =\frac{B_1-2\chi }{%
C_1+B_1\chi -\chi ^2}\left( \chi _r^2+\chi _z^2\right) .  \label{hfiseven}
\end{equation}
Assuming the functional dependence $\chi \left( \psi \right) $ and $\Delta
\psi =0$, the analogues of Eqs (36,37) follow 
\begin{equation}
\psi =\int \frac{d\chi }{C_1+B_1\chi -\chi ^2},  \label{hfieight}
\end{equation}
\begin{equation}
\chi _i=\frac{r^2}f\psi _i,\quad \lambda _r=r\psi _z,\quad \lambda _z=-r\psi
_r.  \label{hfinine}
\end{equation}
The function $f$ should remain positive when the magnetic field is turned
off, hence, $C_1>0$. It can be set to one, as we shall see. The analogue of $%
D$, $D_1=B_1^2+4C_1$ is always positive, leaving only the option $D_1>0$ for
the integral in Eq. (158). This time, however, $\left( B_1-2\chi \right)
^2<D_1$ and one finds instead of Eqs. (42-43) 
\begin{equation}
\chi =\frac{B_1}2+\frac{\sqrt{D_1}}2\tanh \frac{\sqrt{D_1}}2\psi ,\quad f=%
\frac{4r^2}{D_1}\cosh ^2\frac{\sqrt{D_1}}2\psi .  \label{hsixty}
\end{equation}
Since $\cosh x>1$ always, $f\rightarrow \infty $ when $r\rightarrow \infty $
and $\psi _0$ is useless. We put $\psi _0\equiv 0$. Obviously, $f$ does not
fit for an exterior solution except, possibly, in the cylindrical case. When 
$\psi =\psi \left( z\right) $ we have $f\left( r,z\right) $ and already $f$
breaks inheritance. For a regular interior solution at the origin, $\psi $
should compensate the factor $r^2$. Using Eq. (32) with $\phi $ replaced by $%
\lambda $, one obtains after some calculations 
\begin{equation}
k=\ln \frac fr+\tilde k,  \label{hsione}
\end{equation}
\begin{equation}
\tilde k_r=\frac{D_1}4r\left( \psi _r^2-\psi _z^2\right) ,\quad \tilde k_z=%
\frac{D_1}2r\psi _r\psi _z.  \label{hsitwo}
\end{equation}
Here $\tilde k$ satisfies the same equation (38) as $k$ in the Weyl case.
Acceleration is obtained from Eqs. (7,157) 
\begin{equation}
\frac{g_i}{c^2}=\frac{\delta _{\;i}^r}r+\frac{\chi -B_1/2}{1+B_1\chi -\chi ^2%
}\chi _i  \label{hsithree}
\end{equation}
and obviously there is a root gravity term $\sim B_1\sqrt{G}\bar \chi _i/2$.
One can also express $\chi $ in terms of $f$ and $\psi _i$ from Eqs.
(157,159) 
\begin{equation}
\chi =\frac 12\left( \ln \frac f{r^2}\right) _i\psi _i^{-1}+\frac{B_1}2.
\label{hsifour}
\end{equation}
This formula is analogous to Eq. (135) in the electric case.

When the magnetic field is turned off, the line element becomes 
\begin{equation}
ds^2=r^2\left( dx^0\right) ^2-\left( dr^2+dz^2+d\varphi ^2\right) .
\label{hsifive}
\end{equation}
According to one interpretation $r$ and $\varphi $ become cartesian
coordinates like $z$ and this is the Rindler solution representing flat
spacetime in accelerating coordinate system. This means that the static
source will be accelerating when we pass to a non-moving coordinate system
and is not reasonable physically. The second interpretation is to consider
Eq. (165) as the $\nu =1/2$ case of the LC solution (139). There are
well-known problems with the physical interpretation of this metric when $%
\nu >1/2$. Recently a source in the form of a cylindrical shell of
anisotropic fluid was found for $0\leq \nu <\infty $ \cite{heightt}. Out of
the several definitions of mass density per unit length, the most reasonable
seems to be the Israel's one, which gives 
\begin{equation}
m_cc^2=\frac \nu {4\nu ^2-2\nu +1}.  \label{hsisix}
\end{equation}
The mass reaches a maximum exactly for $\nu =1/2$, this being a turning
point. The coordinates $z$ and $\varphi $ switch their meaning for $\nu >1/2$%
. On the other side, the LC solution is flat for $\nu =0,1/2,\infty $. The
general impression is that an infinite line mass field has been imposed upon
a usual Weyl solution, endowing it with some cylindric features. Such metric
with parasitic masses and singularities cannot compete with the genuine Weyl
solutions except for one case.

Let us find the magnetic analogue of the charged cylinder. One has 
\begin{equation}
\bar \psi =H_0\ln \frac r{r_0},\quad \bar \psi _r=\frac{H_0}r,
\label{hsiseven}
\end{equation}
\begin{equation}
f=r^2\Gamma _{+}^2,\quad \tilde k=n^2\ln r,\quad n=\frac{\sqrt{GD_1}H_0}{2c^2%
},  \label{hsieight}
\end{equation}
\begin{equation}
\chi =\frac{\left( \Gamma _{+}\right) _r}{\psi _r\Gamma _{+}}+\frac{B_1}2%
,\quad H_z=r\bar \psi _r=H_0  \label{hsinine}
\end{equation}
and $H_r=0$. Eqs. (142,159) have been used, while $\Gamma _{+}$ is taken
from Eq. (133) with $D\rightarrow D_1$. The physical magnetic field is 
\begin{equation}
H_{\left( z\right) }=F_{\left( r\right) \left( \varphi \right) }=\left(
g_{\varphi \varphi }g_{rr}\right) ^{-1/2}F_{r\varphi }=\frac fre^{-k}\bar 
\chi _r=H_0e^{-k}.  \label{hseventy}
\end{equation}
This solution is similar to the one given by Eqs. (133-135) and has the line
element 
\begin{equation}
ds^2=r^2\Gamma _{+}^2\left( dx^0\right) ^2-r^{2n^2}\Gamma _{+}^2\left(
dr^2+dz^2\right) -\Gamma _{+}^{-2}d\varphi ^2.  \label{hseone}
\end{equation}
It is regular at $r=0$ and may be used as an interior solution only when $%
n=1 $. Then we put $r_0^2=D_1$ to obtain flat spacetime when $H_0=0$ and
find 
\begin{equation}
ds^2=f\left[ \left( dx^0\right) ^2-dr^2-dz^2\right] -f^{-1}r^2d\varphi ^2,
\label{hsetwo}
\end{equation}
\begin{equation}
f=\left( 1+\frac \kappa {32\pi }H_0^2r^2\right) ^2.  \label{hsethree}
\end{equation}

The electric analogue of this solution was discovered by Bonnor in 1953 \cite
{ninet}, while studying solutions with $\phi =\phi \left( z\right) $. Its
magnetic version was found a year later \cite{five}. Later it was obtained
implicitly in Ref. \cite{hninet}. Melvin \cite{six} discussed its properties
again in connection with Wheeler's theory of geons. It is known as Melvin's
magnetic universe, since it can be used globally. Gautreau and Hoffman \cite
{htwenty} gave the more general solution in Eqs. (160-162) and showed that
the Bonnor solutions follow from it. It was also derived through coordinate
modelling \cite{twsix}. Here we have proved that this is the most general
Weyl solution for the system (156). It is not a Weyl solution for the
original system (155), but still has a root gravity term. The general
cylindrically symmetric solution with $H_z$ field is given in Ref. \cite
{three} by Eq. (22.11). However, this is not in the axially-symmetric form
because $g_{rr}\neq g_{zz}$. That is why it differs from Eq. (171), although
Eq. (22.13) coincides with Eqs. (172-173). A combination of LC and
cylindrical electromagnetic solutions has been also studied in the Rainich
formalism \cite{htwone,htwtwo} and in the usual Einstein formalism \cite
{four,heleven}.

One can invoke the ''mass out of charge'' mechanism exactly as in the case
of the charged cylinder. We set $\sqrt{D_1}=r_0^n$ and send $D_1$ to
infinity, keeping $n$ finite. The charge $q$ is replaced by $H_0$. The LC
solution (139) is obtained with $n=1-2\nu $. This time $B_1\chi $ is
preserved and one gets an acceleration formula from Eqs. (163,169) 
\begin{equation}
\frac{g_r}{c^2}\rightarrow \frac 1r-\frac{\sqrt{D_1}H_0}{2r}=\frac{2\nu }r.
\label{hsefour}
\end{equation}
The final result is the same as in Eq. (140), namely the acceleration in a
LC gravitational field.

After this digression, let us return to the exceptional case, Eq. (173).
There are no root gravity terms in this formula and one may think that at
last this is the example of vanishing $B_1$. However, the regularity
condition $n=1$ leads to $B_1=2\sqrt{c^4/G-H_0^2}/H_0$, which, in general,
is not zero. It is worth to trace, using the formulae from this section, how
the acceleration loses its root gravity term, due to the regularity
condition 
\begin{equation}
\frac{g_r}{c^2}=\frac 1r+\left( \chi -\frac{B_1}2\right) \psi _r=\frac 1r%
+\left( \ln \Gamma _{+}\right) _r=\frac{2r\kappa H_0^2}{32\pi +r^2\kappa
H_0^2}.  \label{hsefive}
\end{equation}

We have mentioned that the Weyl cylindrical solution is always singular at
the axis, hence, for an infinitely long solenoid the Bonnor-Melvin solution
should be used. However, this does not seem to describe any physical
situation. In any real situation, starting from the current loop, going
through finite solenoids with open ends and ending with the closed-surface
spheroidal ones, there exist a regular Weyl solution (depending on $z$ or $z$
and $r$) with root gravity term. The infinite solenoid appears to be the
unphysical limit in a series of elongated prolate spheroidal solenoids,
where $g_z\neq 0$ changes into $g_r\neq 0$ and root gravity into usual
gravity.

It should be noted that there is another solution for $\psi $ except Eq.
(167), which makes $f$ in Eq. (160) regular. Separating the variables in the
Laplace equation in cylindrical coordinates, the radial dependence of $\psi $
is given by the Bessel functions $J_0\left( r\right) $,$N_0\left( r\right) $%
. The latter possesses a logarithmic singularity, which can compensate the $%
r^2$ term in $f$.

The transfer of the system (155) into the system (156) has, due to the
Bonnor transformation, its mirror among stationary vacuum solutions. It is
given by Theorem 19.3 from Ref. \cite{three}, based on Ref. \cite{htwthree}
and rediscovered several times \cite{htwfour,htwfive}. The functional
dependence in Eqs. (157,158) translates into the so-called $S\left( A\right) 
$ solutions (\cite{three}, Sec. (20.4) 
\begin{equation}
S\equiv r^2r^{-4u_s}=\omega ^2+B_2\omega +C_2,\quad \psi _s=\int \frac{%
e^{4u_s}d\omega }{r^2},  \label{hsesix}
\end{equation}
where $\Delta \psi _s=0$. They are also known as the Ehlers class \cite
{sefour,htwsix} and were investigated recently \cite{htwseven}. The
elucidation of this connection comes to show that the study of $S\left(
A\right) $ solutions began already in 1953.

\subsection{Line current}

In Maxwell magnetostatics when a current $I$ flows in a straight line
conductor along the axis $z$ with radius $r_1$, it creates a circular
magnetic field in the outside 
\begin{equation}
H_\varphi =\frac{2I}{cr},  \label{hseseven}
\end{equation}
where CGS units are used, the current included. It follows either from a
scalar potential $\psi \left( \varphi \right) $ or from a vector potential
with only a $a_z$ component 
\begin{equation}
\psi =-\frac{2I}c\varphi ,\quad a_z=-\frac{2I}c\ln r.  \label{hseeight}
\end{equation}
In four-dimensional notation $a_z=A_z$. In general relativity, instead of $%
H_\varphi $ defined by Eq. (141) one should use in this case the physical
component $H_{\left( \varphi \right) }$ \cite{thirt} 
\begin{equation}
H_{\left( \varphi \right) }=F_{\left( z\right) \left( r\right) }=\frac 1r%
H_\varphi .  \label{hsenine}
\end{equation}
In flat spacetime it yields the usual definition $\vec H=curl\vec a$. Such
type of field breaks the circularity condition, necessary for the usual
projection formalism of stationary (and static) axisymmetric fields (\cite
{three}, Sec.19.2). Nevertheless, in the cylindrically symmetric case the
solution is known explicitly \cite{htweight} (see also \cite{three}, Eq.
(22.11)) 
\begin{equation}
ds^2=r^{2n^2}\Gamma ^2\left( \left( dx^0\right) ^2-dr^2\right) -r^2\Gamma
^2d\varphi ^2-\Gamma ^{-2}dz^2,  \label{heighty}
\end{equation}
\begin{equation}
\xi =\frac{r\Gamma _r}{a_3\Gamma },\quad \Gamma =a_1r^n+a_2r^{-n},\quad
a_3=4a_1a_2n^2>0,  \label{heione}
\end{equation}
where $\bar \xi =A_z$. The metric of the infinitely long solenoid (171) also
can be written in this form when the changes $A_z\rightarrow A_\varphi $, $%
z\rightarrow \varphi $ are made. Eq. (180) is wrongly quoted in Ref. \cite
{three} as referring to the long solenoid - the Melvin metric, Eq. (22.13),
does not follow from it. One should consult the first edition of Ref. \cite
{three} on this question. After \cite{htweight}, the gravitational field of
a line current was studied also in Refs. \cite{heleven,htwone,htwtwo}.
Combining Eqs. (179,180), one obtains 
\begin{equation}
H_{\left( \varphi \right) }=r^{-n^2}\bar \xi _r.  \label{heitwo}
\end{equation}

In order to generalize Eq. (177) one should use Eq. (10.87) from Ref. \cite
{thirt} (see also \cite{htwtwo}) 
\begin{equation}
\frac{4\pi }cI=\oint F_{\left( z\right) \left( r\right) }\sqrt{g_{\varphi
\varphi }g_{tt}}d\varphi dt  \label{heithree}
\end{equation}
Here $0\leq t\leq 1$ and the circular integral holds for any $r\geq r_1$.
This gives 
\begin{equation}
\bar \xi _r=r^{n^2}H_{\left( \varphi \right) }=\frac{2I}{cr\Gamma ^2}.
\label{heifour}
\end{equation}
The flat spacetime limit follows when $n=0$, $\Gamma =1$. The field is
non-Weyl, but has a root gravity term. Indeed, we can use the analogy with
the long solenoid's equations (157,168) to obtain 
\begin{equation}
\Gamma ^{-2}=1+B_1\xi -\xi ^2,\quad g_{00}=\frac{r^{2n^2}}{1+B_1\xi -\xi ^2}.
\label{heifive}
\end{equation}
Then Eq. (7) gives the analogue of Eq. (163) 
\begin{equation}
\frac{g_r}{c^2}=\frac{n^2}r+\left( \xi -\frac{B_1}2\right) \Gamma ^2\xi _r.
\label{heisix}
\end{equation}
Replacing Eq. (184) into this equation gives for the dominating root gravity
term 
\begin{equation}
g_r=-0.1B_1\sqrt{G}\frac Ir,  \label{heiseven}
\end{equation}
where $I$ is in $Amps$. The metric (180) is singular at the axis, but anyway
we use it as an exterior. Therefore, the regularity mechanism leading to the
Melvin magnetic universe is not needed here. Unfortunately, there are no
hints for the scale of $B_1$. Assuming its typical value $2$, finally yields 
\begin{equation}
g_r=-5.16\times 10^{-5}\frac I{r_1}.  \label{heieight}
\end{equation}
For a ''static'' lightning $I=10^5A$, $r_1=10cm$, $g_l=0.516cm/s^2$. The
magnitude of $g_e$ is obtained for, let us say, $r_1=1cm$, $I_e=1.9\times
10^7A$. The gravitational force is radial, like in the long solenoid
example, does not follow the magnetic lines, but is perpendicular to them.

\section{Conclusions}

When the derivations, proofs and other technicalities in this paper are
omitted, the summary of the results remains, divided into 19 points. Some of
them are not new and the corresponding references are cited in these cases.

1) The gravitational acceleration at rest in Weyl-Majumdar-Papapetrou fields
has a root gravity term, proportional to $c^2\sqrt{\kappa }=\sqrt{G}$, which
is $10^{23}$ times bigger than the usual perturbative coefficient $c^2\kappa 
$. It is linear in the EM-fields, while the perturbative term is quadratic.
Sizeable gravitational force exists (see Eq. (53)) although the metric is
very close to the flat one (Maxwellian limit). Its explicit form determines
up to a sign the important constant $B$. For its typical value $B=2$ the
Earth's acceleration is obtained in electric fields of order $10^9V/cm$ and
in magnetic fields of about $380T$. One can change the direction of $g_i$ by
changing the direction of $E_i$ or $H_i$ and confine $g_i$ to a finite
volume by confining the EM-fields.

2) In WMP fields the gravitational potential depends directly on the
four-potential of the EM-fields. This is not amazing - such potentials are
known to be important in certain quantum effects. One should not make gauge
transformations because they spoil the asymptotic behaviour of the metric or
the initial condition - when the EM-field is turned off, flat spacetime
should result. Coordinate transformations are formally equivalent to
modification of the symmetry of the charge (current) carrying surface and to
a change in the physical set-up.

3) The energy-momentum tensor (in particular, its energy component $T_{00}$)
induces a change in the Ricci tensor $\sim \kappa $ according to the
Einstein equations (3). This leads to changes in the metric and its
acceleration, which can be $\sim \sqrt{\kappa }$ and may contain no monopole
term. Creating artificial gravity that is localized in space and has no
long-distance mass terms seems energetically more favourable.

4) In axially-symmetric systems Weyl fields provide regular exterior and
interior solutions to any distribution of charges or currents on a closed
surface. They are determined by a master-potential $\psi $ satisfying the
Laplace equation. When the metric depends on just one coordinate (three
commuting Killing vectors are present) the Weyl solution becomes the most
general one. In truly axisymmetric cases it presumably determines the pure
electromagnetic effect on gravity, while the solutions of the Ernst equation
include hidden mass sources. The constant $B$ (taken positive for
definiteness) divides the Weyl solutions into three classes, according to $%
B=2,$ $B>2$, $B<2$. \cite{ninet}. Among them $B=2$ is privileged, being the
simplest (conformastatic spacetimes). When $B\neq 2$, parasitic masses
appear, either on the surface through the junction formalism, or on the
axis. In some cases the metric does not inherit the symmetry of the
EM-source.

5) The gravitational force at rest induced by electric and magnetic fields
is the same \cite{fift,sixt,sevent} unlike the Lorentz force, acting upon
charged particles. The surface sources determine the master potential and
not $A_\mu $ \cite{five,ninet}.

6) There is a ''mass out of charge'' mechanism, which allows to obtain
solutions with mass and charge from the Weyl solutions. It clearly indicates
the part of mass which is of electromagnetic origin. Eq. (14) still holds
and root gravity term remains, but $B$ is affected by the mass. Such
solutions can incorporate the mass of the charged surface, which is always
present in practice.

7) In the general static case a harmonic master potential for WMP fields
appears only when $B=2$ \cite{eleven,twelve}. Point or line sources have
singularities, hence, one should use closed surface sources (shells).
Another alternative is to use volumes of charged dust or perfect fluid,
where the functional dependence $f\left( \phi \right) $ appears naturally as
an equilibrium condition, but is arbitrary \cite{fitwo}. For charged dust a
harmonic potential can always be introduced \cite{fitwo}, but the
equilibrium is unstable \cite{sithree,sifour,sifive,sisix}. It is more
realistic to charge conductive surfaces by applying potentials or pass
currents through coils wound around them, relying on the motion of free
electrons in metals. One can also create powerful fields by separating the
positive and negative charged ions.

8) The wealth of solutions to the stationary vacuum Ernst equation is
preserved completely for the static electro(magneto)vacs due to the Bonnor
transformation \cite{foone,sinine}. Usual Weyl solutions correspond to the
Papapetrou stationary solutions, while Weyl solutions to the transformed
system of equations (156) correspond to the $S\left( A\right) $ solutions.
They consist of only one class $D>0$ and although non-Weyl, possess root
gravity terms. The case $B=2$ is special also for stationary solutions.

9) In order to construct global solutions around the charged shells, the
fulfilment of the junction conditions is required. They have some subtleties
in the Weyl case. Eq. (14) interrelates the Einstein and the Maxwell
equations (31) and a jump in $T_{\mu \nu }$ results due to the jump in $\psi 
$, caused by the charges or the currents. The term in $S_{ab}$, containing $%
u_i$ is absent and a mass surface layer appears only when $k_i$ is
non-trivial. The acceleration is caused by the EM-field and not by often
unrealistic fluid sources, trying to duplicate this effect.

10) The pure electric plane-symmetric effect on the metric is described by
the McVittie solution, which is a Weyl solution of class $B=2$. When $B\neq
2 $ the Weyl metric is not plane-symmetric (non-inheritance). Solutions with
mass and charge also contain a root gravity term. There is a coordinate
transformation between them and a Weyl axisymmetric field, which changes the
plane into a paraboloid.

11) A usual freely hanging capacitor is an excellent tool for testing root
gravity. Constant acceleration is induced in its dielectric, Eq. (89).
Taking maximum potential difference, minimum distance between the plates and
a material with very big dielectric constant (ferroelectric) one can obtain $%
g_z=5.2g_e$. This force is trying to move the capacitor in the $z$
direction. It is smaller than the force of electric attraction between the
plates.

12) The pure electric spherically-symmetric effect on the metric is
described by the charged Curzon solution, Eq. (108), which is related by a
simple transformation to the critically (extremely) charged RN solution.
This is a Weyl solution with $B=2$. When $B\neq 2$ non-inheritance follows.
The same is true for prolate and oblate spheroidal solutions. They are
singular at the axis, so one can use them as exteriors to a charged
spheroidal shell with Minkowski interior. Their gravity becomes repulsive in
the whole outside region for one sign of the charge. Typically, the
repulsive $g=0.5cm/s^2$. There is a coordinate transformation between RN
fields with mass and charge and Weyl fields outside spheroidal charged
shells.

13) An exact Weyl solution exists with practically constant gravitational
force inside a charged sphere, which quickly decreases outside, according to
a dipole law. Unfortunately, with potentials mankind can create at present,
the force inside this microgravity chamber is very weak, although still
detectable.

14) The pure electric cylindrically-symmetric effect on the metric is
described by Eq. (131), which is regular outside the shell. In non-regular
form it was first given by Bonnor \cite{ninet} and is the electric
counterpart of the limiting Van Stockum stationary solution. Once again, $%
B=2 $. When $B\neq 2$, a mass surface layer appears on the shell, inducing
the second gravitational potential $k$. The ''mass out of charge'' mechanism
leads to the vacuum LC solution.

15) The magnetic field of a current loop induces gravitational force, Eq.
(147). The Earth's acceleration is reached when the current $I=6\times 10^8A$%
. A thick loop (fat solenoid) requires for this goal current density of $%
3.8\times 10^4A/cm^2$. Presumably, a ferromagnetic disc with a current strip
around it is the magnetic analogue of the moving capacitor. The ambiguity in
the sign of $B$ seems to be necessary to make the metric unique and
continuous across the plane of the loop.

16) A spheroidal solenoid represents the magnetic analogue of the electric
microgravity chamber. The uniform gravitational force inside equals that of
Earth when $J=4.5\times 10^6A/cm$. Outside the fields quickly vanish, being
dipole in character.

17) The Weyl magnetic solutions of the transformed system of equations (156)
are magnetovac mirrors of the stationary $S\left( A\right) $ solutions. They
have singularities due to parasitic infinite line masses, consist of only
one class and possess root gravity terms. In the cylindrically-symmetric
case the metric is similar to the charged cylinder, but it is regular at the
axis when a special condition is satisfied (Bonnor-Melvin solution, also
used in the interior of a infinitely long solenoid \cite{four}). The root
gravity term is sacrificed in favour of regularity, resulting in an
acceleration $\sim \kappa H_0^2$, which is negligible for realistic magnetic
fields on Earth. The infinite solenoid seems to be unphysical limit, because
in any finite solenoid there exists a regular Weyl solution with root
gravity term.

18) The cylindrical gravitational field of a line current breaks the
circularity condition of the axial symmetry formalism, but nevertheless can
be found explicitly \cite{three,htweight}. It is similar to the field of a
charged cylinder or to the field of the magnetic solution from the previous
point. It is non-Weyl, but still has a root gravity term. If $B=2$ a current
of magnitude $1.9\times 10^7A$ creates a radial gravitational force, equal
to the Earth's one at $r=1cm$.

19) Some statements in the literature have been corrected. The Weyl case $%
D>0 $ does not follow simply by analytic continuation from $D<0$. A number
of misunderstandings in the cylindrically-symmetric Einstein-Maxwell fields
in Ref. \cite{three} has been clarified.

\section{Discussion}

It is hardly believable that such an important property of WMP fields as
root gravity has been overlooked in the past. If we consider the work of
Reissner as implicitly pioneering, WMP solutions date back to 1916, the year
of the Schwarzschild solution. In the past 88 years the subject has been
explored pretty thoroughly as can be seen from the number of references.
Only in the first half of the previous century root gravity could have been
discovered 8 times: by Reissner, 1916 \cite{hthree}, Weyl, 1917 \cite{fourt}%
, Nordstr\"om, 1918 \cite{hfour}, Kar, 1926 \cite{eifive}, McVittie, 1929 
\cite{nifive}, Mukherji, 1938 \cite{htwelve}, Papapetrou, 1947 \cite{twelve}
and Majumdar, 1947 \cite{eleven}. There were plenty of opportunities later
too. In the same time certain spherically-symmetric perfect fluid solutions
have been rediscovered up to 7 times.

One reason is the wide-spread use of relativistic units - already Weyl
worked in them. The only papers on the subject (except the present one)
written in non-relativistic units are by Ehlers \cite{thnine,forty}.
Unfortunately, they are in German and are extremely rarely cited. When $%
G=c=1 $, $\sqrt{\kappa /8\pi }=\kappa /8\pi =1$ and there is no difference
between root and usual gravity. When $8\pi G=c=1$ then $\kappa =1$ and $%
\sqrt{\kappa /8\pi }=0.2$ while $\kappa /8\pi =0.04$. The difference is just
an order of magnitude.

Another reason is that the easier solutions with plane, spherical,
spheroidal or cylindrical symmetry were studied directly and were almost
never treated as subcases of the axisymmetric solutions. This cut their ties
to Weyl solutions from the very beginning. The RN solution was shown to
belong to the Weyl class only in 1972 \cite{foseven}.

A third reason is the shaky position of the linear term in Eq. (14), which
can be eliminated in principle by a gauge transformation. In many papers
disappearance has been its fate. We have argued that physically this is not
possible and the typical value of $B$ is $2$.

A fourth reason is the underestimation of exact solutions in favour of
approximation schemes and numerical techniques. Root gravity escapes
undetected by these methods. In fact, it sets a new scale $\sqrt{\kappa }$
or $\sqrt{G}$ and perturbations should be done around the exact Weyl
solutions.

More generally, WMP solutions have never been in the main trend of
development of general relativity. Black holes, perfect fluid star models
and cosmological solutions write its official history. The fact that the
unique charged black hole is a WMP solution pushed their investigation
towards multiple-charged-black holes \cite
{eleven,twelve,thirt,twenty,fofive,fosix,foseven,foeight,fonine,fifty},
particle trajectories in these fields \cite{htwnine} or even WMP-based
wormholes \cite{hthirty}. A similar astrophysical string rings in charged
dust clouds (Bonnor stars) and their non-singular interpolation to black
holes \cite{fifour,fifive,fisix,fiseven,fieight,hthone,hthtwo}. Axisymmetric
solutions have been flooded with generation techniques for the Ernst
equation. Papers on WMP solutions often appeared in local, old or unpopular
journals, unavailable to the scientific community. Even the Weyl seminal
paper still remains not translated into English.

A sixth reason concerns the motivation of the first researchers. Weyl
himself was much more interested in his conformal theory of gravity, trying
to unify gravitation and electromagnetism. He never returned back to the
Weyl fields. Ironically, McVittie presented his solution as an example to
test one of the unified theories of Einstein. The father of relativity used
to complain about the difficulties of exact solutions and found only four of
them (Einstein static universe, cylindrical waves, spherical cloud of
counter-rotating particles and the Einstein-Strauss vacuole). He
concentrated his energy on the geometrical unification of the two
fundamental long-range interactions. According to him, the l.h.s. of Eq. (3)
was made of marble, while the r.h.s. was made of wood. His goal was to
replace wood by marble.

In this paper we share another view, closer to the Rainich ''already unified
theory''. We are interested in the electromagnetic effects upon gravity. It
is not necessary to use his formalism, which includes products of Ricci
tensors and is similar in complexity to the Gauss-Bonnet terms in the string
effective action \cite{hththree}. If $F_{\mu \nu }$ is wooden, let us do
some alchemy, let us obtain marble from wood, gold from lead. The point is
that electromagnetism has been mastered for centuries on Earth and we know
how to create strong EM-fields. As for gravity, we only know how to watch
this force in action around the Universe.

Our approach is not to explain but to construct and parallels the efforts to
build wormholes, time machines and warp drives. We use, however, fields
which are common and satisfy all energy conditions and not exotic matter or
the Kasimir effect. Second, we try to induce gravitational force without
altering very much the metric. This is possible thanks to the 20 orders of
magnitude collected in $c^2/2$ in Eq. (17). We have demonstrated that except
the Newtonian limit, general relativity possesses also a Maxwellian limit.
Root gravity is one way to do this, but there are others too. The next step
should be to change considerably the metric of spacetime and invoke typical
Einsteinian effects. It seems more difficult and expensive, but perhaps
among the 450 line elements given in Ref. \cite{three} or in the 1822
references quoted there, there waits the necessary solution.

Our tool is well-known electromagnetism. It is hard to imagine that the
strong or the weak interactions, which deal with quark confinement and
particle decays, can be used to manipulate gravity. At least root terms are
not possible. QCD and QFD interact through non-Abelian gauge fields and in
the r.h.s. of Eq. (1) there will be cubic and quartic terms in $A_\mu $.
This makes impossible to swallow the Einstein constant as this was done in
Eq. (12).

It seems that the first who studied experimentally the interaction between
gravity and electromagnetism was Faraday \cite{hthfour}. He dropped or
lifted heavy bodies inside solenoids or measured their charge after such
movements. He expected that gravity would induce currents or charges, but
the results were negative. His last paper, which described these futile
experiments, was written in 1860 and rejected by Phil. Transactions. Two of
the decisive experiments confirming general relativity included the effects
of static masses on null EM-fields (light bending around the sun and the
redshift of photons in a gravitational field). It is time to do the
opposite. It is the electromagnetic effect upon gravity that is much more
interesting. Unfortunately, the main device for strong magnetic fields - the
usual solenoid, is often considered infinite for simplicity, which results
in a solution without root gravity terms \cite{four,eight,nine,ten}. We have
given here several laboratory set-ups where root gravity can be detected and
general relativity tested once more, this time in its highly non-linear
part. The most promising seems to be the well-known capacitor and its
magnetic analogue. The effect has not been discovered yet, because
capacitors are used in low-voltage circuits and they are first firmly fixed
and charged afterwards. Small root gravity effects should be present also in
today's solenoids with strong magnetic fields. The acceleration is masked by
that of the Earth, being a fraction of it.


\begin{thebibliography}{1}
\bibitem{one}  Rainich G Y 1925 {\it Trans. Amer. Math. Soc.} {\bf 27}, 106

\bibitem{two}  Misner C W and Wheeler J A 1957 {\it Ann. Phys.} {\bf 2}, 525

\bibitem{three}  Stephani H, Kramer D, MacCallum M A H, Hoenselaers C and
Herlt E 2003 {\it Exact Solutions of Einstein's Field Equations }(2nd
edition, Cambridge: Cambridge University Press).

\bibitem{four}  Ivanov B V 1994 {\it Mod. Phys. Lett. A} {\bf 9}, 1627

\bibitem{five}  Bonnor W B 1954 {\it Proc. Phys. Soc. A} {\bf 67}, 225

\bibitem{six}  Melvin M A 1964 {\it Phys. Lett.} {\bf 8}, 65

\bibitem{seven}  Levi-Civita T 1917 {\it Rend. Acc. Lincei} {\bf 26}, 307

\bibitem{eight}  Ivanov B V 1997 {\it Mod. Phys. Lett. A} {\bf 12}, 285

\bibitem{nine}  Ivanov B V 1994 {\it Class. Quantum Grav.} {\bf 11}, 1359

\bibitem{ten}  Ivanov B V 1996 {\it Phys. Lett. A} {\bf 210}, 255

\bibitem{eleven}  Majumdar S D 1947 {\it Phys. Rev.} {\bf 72}, 390

\bibitem{twelve}  Papapetrou A 1947 {\it Proc. R. Irish. Acad.} A {\bf 51},
191

\bibitem{thirt}  Synge J L 1960 {\it Relativity: the General Theory}
(Amsterdam: North-Holland)

\bibitem{fourt}  Weyl H 1917 {\it Ann. Phys. (Germany)} {\bf 54}, 117

\bibitem{fift}  Perj\'es Z 1968 {\it Nuovo Cim. B} {\bf 55}, 600

\bibitem{sixt}  Perj\'es Z 1968 {\it Acta Phys. Acad. Sci. Hung.} {\bf 25},
393

\bibitem{sevent}  Das A 1979 {\it J. Math. Phys.} {\bf 20}, 740

\bibitem{eightt}  Tauber G 1957 {\it Can. J. Phys.} {\bf 35}, 477

\bibitem{ninet}  Bonnor W B 1953 {\it Proc. Phys. Soc. A} {\bf 66}, 145

\bibitem{twenty}  Hartle J B and Hawking S W 1972 {\it Comm. Math. Phys.} 
{\bf 26}, 87

\bibitem{twone}  Bonnor W B 1960 {\it Proc. Phys. Soc. A} {\bf 76}, 891

\bibitem{twtwo}  Whittaker E T 1935 {\it Proc. R. Soc. A} {\bf 149}, 384

\bibitem{twthree}  Bonnor W B 1992 {\it Gen. Rel. Grav.} {\bf 24}, 551

\bibitem{twfour}  Carminati J and Cooperstock F I 1982 {\it Phys. Lett. A} 
{\bf 90}, 349

\bibitem{twfive}  Carminati J and Cooperstock F I 1983 {\it Phys. Lett. A} 
{\bf 94}, 27

\bibitem{twsix}  Carminati J and Cooperstock F I 1983 {\it J. Phys. A} {\bf %
16}, 3867

\bibitem{twseven}  Ernst F J 1968 {\it Phys. Rev.} {\bf 168}, 1415

\bibitem{tweight}  Ansorg M, Kleinw\"achter A, Meinel R and Neugebauer G
2002 {\it Phys. Rev. D} {\bf 65}, 044006

\bibitem{twnine}  Whittaker E T 1903 {\it Math. Ann.} {\bf 57}, 333

\bibitem{thirty}  Whittaker E T and Watson G N 1927 {\it A Course of Modern
Analysis }(Cambridge: Cambridge University Press).

\bibitem{thone}  Waylen P C 1982 {\it Proc. R. Soc. A} {\bf 382}, 467

\bibitem{thtwo}  Guilfoyle B S 1999 {\it J. Math. Phys.} {\bf 40}, 2032

\bibitem{ththree}  Xulu S S 2003 {\it hep-th} / 0308070.

\bibitem{thfour}  Letelier P S 1999 {\it Phys. Rev. D} {\bf 60},104042

\bibitem{thfive}  Katz J, Bi\v c\'ak J and Lynden-Bell D 1999 {\it Class.
Quantum Grav.} {\bf 16}, 4023

\bibitem{thsix}  Papapetrou A 1966 {\it Ann. Inst. H. Poincar\'e A}{\bf 4},
83

\bibitem{thseven}  Horsk\'y J and Mitskievitch N V 1989 {\it Czech. J. Phys.
B }{\bf 39}, 957

\bibitem{theight}  Richterek L, Novotn\'y J and Horsk\'y J 2002 {\it Czech.
J. Phys. B }{\bf 52}, 1021

\bibitem{thnine}  Ehlers J 1955 {\it Z. Phys. }{\bf 140}, 394

\bibitem{forty}  Ehlers J 1955 {\it Z. Phys. }{\bf 143}, 239

\bibitem{foone}  Bonnor W B 1961 {\it Z. Phys. }{\bf 161}, 439

\bibitem{fotwo}  Teixeira A F da F, Wolk I and Som M M 1976 {\it J. Phys. A }%
{\bf 9}, 53

\bibitem{fothree}  Som M M, Santos N O and Teixeira A F da F 1977 {\it Phys.
Rev. D }{\bf 16}, 2417

\bibitem{fofour}  Barbachoux C, Gariel J, Marcilhacy G and Santos N O 2002 
{\it Int. J. Mod. Phys. D }{\bf 11}, 1255

\bibitem{fofive}  Chru\'sciel P T and Nadirashvili N S 1995 {\it Class.
Quantum Grav. }{\bf 12}, L17

\bibitem{fosix}  Heusler M 1997 {\it Class. Quantum Grav. }{\bf 14}, L129

\bibitem{foseven}  Gautreau R, Hoffman R B and Armenti A 1972 {\it Nuovo
Cim. B }{\bf 7}, 71

\bibitem{foeight}  Kastor D and Traschen J 1993 {\it Phys. Rev. D }{\bf 47},
537

\bibitem{fonine}  Brill D R, Horowitz G T, Kastor D and Traschen J 1994 {\it %
Phys. Rev. D }{\bf 49}, 840

\bibitem{fifty}  Bonnor W B 2000 {\it Class. Quantum Grav. }{\bf 17}, 3935

\bibitem{fione}  Das A 1962 {\it Proc. R. Soc. A }{\bf 267}, 1

\bibitem{fitwo}  Bonnor W B 1980 {\it Gen. Rel. Grav. }{\bf 12}, 453

\bibitem{fithree}  De U K and Raychaudhuri A K 1968 {\it Proc. R. Soc. A }%
{\bf 303}, 97

\bibitem{fifour}  Bonner W B 1965 {\it Month. Not. R. Astr. Soc. }{\bf 129},
443

\bibitem{fifive}  Bonnor W B and Wickramasuriya 1972 {\it Int. J. Theor.
Phys. }{\bf 5}, 371

\bibitem{fisix}  Bonnor W B and Wickramasuriya 1975 {\it Month. Not. R.
Astr. Soc. }{\bf 170}, 643

\bibitem{fiseven}  Bonnor W B 1998 {\it Class. Quantum Grav. }{\bf 15}, 351

\bibitem{fieight}  Bonnor W B 1999 {\it Class. Quantum Grav. }{\bf 16}, 4125

\bibitem{finine}  G\"urses M 1998 {\it Phys. Rev. D }{\bf 58}, 044001

\bibitem{sixty}  Varela V 2003 {\it Gen. Rel. Grav. }{\bf 35}, 1815

\bibitem{sione}  G\"urses M 1998 {\it gr-qc / 9806038}

\bibitem{sitwo}  Ardavan H and Hossein Partovi M 1977 {\it Phys. Rev. D }%
{\bf 16}, 1664

\bibitem{sithree}  Gautreau R and Hoffman R B 1973 {\it Nuovo Cim. B }{\bf 16%
}, 162

\bibitem{sifour}  Ida D 2000 {\it Prog. Theor. Phys. }{\bf 103}, 573

\bibitem{sifive}  Gautreau R 1985 {\it Phys. Rev. D }{\bf 31}, 1860

\bibitem{sisix}  Guilfoyle B S 1999 {\it Gen. Rel. Grav. }{\bf 31}, 1645

\bibitem{siseven}  Ivanov B V 2002 {\it Phys. Rev. D }{\bf 65}, 104001

\bibitem{sieight}  Harrison B K 1968 {\it J. Math. Phys. }{\bf 9}, 1744

\bibitem{sinine}  Fischer E 1979 {\it J. Math. Phys. }{\bf 20}, 2547

\bibitem{seventy}  Papapetrou A 1953 {\it Ann. Phys. (Germany) }{\bf 12}, 309

\bibitem{seone}  Islam J N 1985 {\it Rotating Fields in General Relativity }%
(Cambridge: Cambridge University Press)

\bibitem{setwo}  Perj\'es Z 1969 {\it Comm. Math. Phys. }{\bf 12}, 275

\bibitem{sethree}  Perj\'es Z 1974 {\it Int. J. Theor. Phys. }{\bf 10}, 217

\bibitem{sefour}  Kloster S, Som M M and Das A 1974 {\it J. Math. Phys. }%
{\bf 15}, 1096

\bibitem{sefive}  Hoenselaers C 1978 {\it Prog. Theor. Phys. }{\bf 60}, 158

\bibitem{sesix}  Zenk L G and Das A 1978 {\it J. Math. Phys. }{\bf 19}, 535

\bibitem{seseven}  K\'ota J and Perj\'es Z 1972 {\it J. Math. Phys. }{\bf 13}%
, 1695

\bibitem{seeight}  Luk\'acs B, Perj\'es Z, Sebesty\'en \'A and Sparling G A
1983 {\it Gen. Rel. Grav. }{\bf 15}, 511

\bibitem{senine}  Perj\'es Z 1986 {\it Gen. Rel. Grav. }{\bf 18}, 511, 531

\bibitem{eighty}  Perj\'es Z 1988 {\it Acta Phys. Hung. }{\bf 63}, 89

\bibitem{eione}  Rayner C B 1959 {\it C. R. Acad. Sci. Paris }{\bf 249}, 1614

\bibitem{eitwo}  Fackerell E D and Kerr R P 1991 {\it Gen. Rel. Grav. }{\bf %
23}, 861

\bibitem{eithree}  Phan D 1991 {\it Gen. Rel. Grav. }{\bf 23}, 269

\bibitem{eifour}  Cosgrove C M 1978 {\it J. Phys. A }{\bf 11}, 2389, 2405

\bibitem{eifive}  Kar S C 1926 {\it Physik. Zeitschr. }{\bf 27}, 208

\bibitem{eisix}  Li J and Liang C 1989 {\it J. Math. Phys. }{\bf 30}, 2915

\bibitem{eiseven}  Kuang Z, Li J and Liang C 1987 {\it Gen. Rel. Grav. }{\bf %
19}, 345

\bibitem{eieight}  Israel W 1966 {\it Nuovo Cim. B }{\bf 44}, 1

\bibitem{einine}  Gr\o n \O\ 1999 {\it Nuovo Cim. B} {\bf 114}, 881

\bibitem{ninety}  Srinivasa Rao K N and Gopala Rao A V 1980 {\it J. Math.
Phys. }{\bf 21}, 2261

\bibitem{nione}  Banerjee A 1970 {\it J. Math. Phys. }{\bf 11}, 51

\bibitem{nitwo}  Letelier P S and Tabensky R R 1974 {\it J. Math. Phys. }%
{\bf 15}, 594

\bibitem{nithree}  Amundsen P A and Gr\o n \O\ 1983 {\it Phys. Rev. D }{\bf %
27}, 1731

\bibitem{nifour}  Das A 1971 {\it J. Math. Phys. }{\bf 12}, 1136

\bibitem{nifive}  McVittie G C 1929 {\it Proc. R. Soc. A }{\bf 124}, 366

\bibitem{nisix}  Patnaick S 1970 {\it Proc. Camb. Phil. Soc. }{\bf 67}, 127

\bibitem{niseven}  Taub A H 1951 {\it Ann. Math. }{\bf 53}, 472

\bibitem{nieight}  Cooperstock F I and de la Cruz V 1978 {\it Phys. Lett. A }%
{\bf 65}, 393

\bibitem{ninine}  Cooperstock F I and de la Cruz V 1979 {\it Gen. Rel. Grav. 
}{\bf 8}, 681

\bibitem{hundred}  Curzon H E J 1924 {\it Proc. Lond. Math. Soc. }{\bf 23},
477

\bibitem{hone}  Bondi H 1957 {\it Rev. Mod. Phys. }{\bf 29}, 423

\bibitem{htwo}  Israel W and Khan K A 1964 {\it Nuovo Cim. }{\bf 33}, 331

\bibitem{hthree}  Reissner H 1916 {\it Ann. Phys. (Germany) }{\bf 50}, 106

\bibitem{hfour}  Nordstr\"om G 1918 {\it Proc. Kon. Ned. Akad. Wet. }{\bf 20}%
, 1238

\bibitem{hfive}  Longo C 1918 {\it Nuovo Cim. }{\bf 15}, 191

\bibitem{hsix}  Cohen J M and Gautreau R 1979 {\it Phys. Rev. D }{\bf 19},
2273

\bibitem{hseven}  Bonnor W B 1966 {\it Z. Phys. }{\bf 190}, 444

\bibitem{height}  Misra R M, Pandey D B, Srivastava D C and Tripathi S N
1973 {\it Phys. Rev. D }{\bf 7}, 1587

\bibitem{hnine}  Lynden-Bell D, Bi\v c\'ak J and Katz J 1999 {\it Ann. Phys. 
}{\bf 271}, 1

\bibitem{hten}  Raychaudhuri A K 1960 {\it Ann. Phys. }{\bf 11}, 501

\bibitem{heleven}  Miguelote A Y, da Silva M F A, Wang A and Santos N O 2001 
{\it Class. Quantum Grav. }{\bf 18}, 4569

\bibitem{htwelve}  Mukherji B C 1938 {\it Bull. Cal. Math. Soc. }{\bf 30}, 95

\bibitem{hthirt}  Safko J L 1970 {\it Ann. Phys. }{\bf 58}, 322

\bibitem{hfourt}  Van Stockum W J 1937 {\it Proc. R. Soc. (Edinburgh) }{\bf %
57}, 135

\bibitem{hfift}  Landau L D and Lifshitz E M 1973 {\it The Classical Theory
of Fields }(Moscow: Nauka)

\bibitem{hsixt}  Montgomery D B 1969 {\it Solenoid Magnet Design }%
(Wiley-Interscience)

\bibitem{hsevent}  Wilson M N 1983 {\it Superconducting Magnets (}Oxford:
Clarendon Press)

\bibitem{heightt}  Herrera L, Santos N O, Teixeira A F F and Wang A Z 2001 
{\it Class. Quantum Grav. }{\bf 18}, 3847

\bibitem{hninet}  Misra M and Radhakrishna L 1962 {\it Proc. Nat. Inst. Sci.
India A }{\bf 28}, 632

\bibitem{htwenty}  Gautreau R and Hoffman R B 1970 {\it Phys. Rev. D }{\bf 2}%
, 271

\bibitem{htwone}  Safko J L and Witten L 1971 {\it J. Math. Phys. }{\bf 12},
257

\bibitem{htwtwo}  Safko J L and Witten L 1972 {\it Phys. Rev. D }{\bf 5}, 293

\bibitem{htwthree}  Kramer D and Neugebauer G 1968 {\it Comm. Math. Phys. }%
{\bf 10}, 132

\bibitem{htwfour}  Sackfield A 1975 {\it J. Phys. A }{\bf 8}, 506

\bibitem{htwfive}  Catenacci R and Alonso J D 1976 {\it J. Math. Phys. }{\bf %
17}, 2232

\bibitem{htwsix}  Ehlers J 1965 {\it in Proc. Int. Conf. on Relat. Theories
of Gravitation }(London)

\bibitem{htwseven}  Gariel J, Marcilhacy G and Santos N O 2003 {\it J. Math.
Phys. }{\bf 44}, 1679

\bibitem{htweight}  Witten L {\it in Gravitation: An Introduction to Current
Research, ed. by L.Witten }(New York: J.Wiley \& Sons)

\bibitem{htwnine}  Yurtsever U 1995 {\it Phys. Rev. D }{\bf 52}, 3176

\bibitem{hthirty}  Schein F and Aichelburg P C 1996 {\it Phys. Rev. Lett. }%
{\bf 77}, 4130

\bibitem{hthone}  Lemos J P S and Weinberg E J 2004 {\it Phys. Rev. D }{\bf %
69}, 104004

\bibitem{hthtwo}  Kleber A, Lemos J P S and Zanchin V T 2004 {\it gr-qc /
0406053}

\bibitem{hththree}  Ivanov B V 1987 {\it Phys. Lett. B }{\bf 198}, 438

\bibitem{hthfour}  Williams L P 1965 {\it Michael Faraday} (London: Chapman
and Hall)
\end{thebibliography}
\end{document}